\documentclass[aps,nofootinbib,notitlepage,showpacs,longbibliography]{revtex4-1}

\usepackage[CJKbookmarks, pdftex, bookmarksnumbered, bookmarksopen, colorlinks, citecolor=magenta, linkcolor=blue]{hyperref}

\usepackage{amstext,amsmath,amssymb,amsfonts,bbm}
\usepackage[latin1]{inputenc}
\usepackage{graphicx}
\usepackage{epstopdf}
\usepackage{amsthm}
\usepackage{tocvsec2}
\usepackage{enumerate}
\usepackage{subfigure}
\usepackage{color}

\usepackage{multirow} \usepackage{rotating}

\def\beq{\begin{equation}}
\def\be{\begin{equation}}
\def\ee{\end{equation}}
\def\bes{\begin{eqnarray}}
\def\ees{\end{eqnarray}}

\begin{document}

\title{(3+1)-Formulation for Gravity with Torsion and Non-Metricity II: The
Hypermomentum Equation}

\author{Seramika Ariwahjoedi$^{1}$, Agus Suroso$^{2,3}$, Freddy P.
Zen$^{2,3}$\vspace{1mm}
 }

\affiliation{$^{1}$Research Center for Physics, Indonesian Institute of Sciences (LIPI), Serpong PUSPIPTEK Area, Tangerang 15310, Indonesia.\\
 $^{2}$Theoretical Physics Laboratory, THEPI Division, Institut Teknologi
Bandung, Jl. Ganesha 10 Bandung 40132, West Java, Indonesia.\\
 $^{3}$Indonesia Center for Theoretical and Mathematical Physics
(ICTMP), Indonesia.}

\begin{abstract}
\noindent In this article, we consider a special case of Metric-Affine
$f(\mathcal{R})$-gravity for $f(\mathcal{R})=\mathcal{R}$, i.e.
the Metric-Affine General Relativity (MAGR). As a companion to the
first article in the series, we perform the (3+1) decomposition to
the hypermomentum equation, obtained from the minimization of the
MAGR action $\mathcal{S}\left[g,\omega\right]$ with respect to the
connection $\omega$. Moreover, we show that the hypermomentum tensor
$\mathcal{H}$ could be constructed completely from 10 hypersurfaces
variables that arise from its dilation, shear, and rotational (spin)
parts. The (3+1) hypermomentum equations consists of 1 scalar, 3 vector,
3 matrix, and 1 tensor equation of order -$\tbinom{2}{1}$. Together
with the (3+1) decomposition of the traceless torsion constraint,
consisting of 1 scalar and 1 vector equation, we obtain 10 hypersurface
equations, which are the main result in this article. Finally, we
consider some special cases of MAGR, namely, the zero hypermomentum,
metric, and torsionless cases. For vanishing hypermomentum, we could
retrieve the metric compatibility and torsionless condition in the
(3+1) framework, hence forcing the affine connection to be Levi-Civita
as in the standard General Relativity.
\end{abstract}
\maketitle

\section{Introduction}

The theories of modified gravity have gained attention in these few
decades \cite{Oikonomou,clifton,CANTATA}. Attempts to modify gravity
arise from the interest to explain exotic phenomenologies such as
dark matter and dark energy, as well as to solve the problem of quantum
gravity. One of the branches of these modified theories is the Metric-Affine
Gravity (MAG) \cite{Hehl,Smalley2,Smalley,Vitagliano,Iosifidis}.
The theory differs from the standard General Relativity (GR) following
the generalization of three objects. The first generalization is on
the action integral \cite{Vitagliano,Iosifidis}. While the standard
GR could be derived from the Einstein-Hilbert action $\mathcal{S}=\int\mathcal{R}\mathrm{vol}$
(with $\mathcal{R}$ is the Ricci scalar as a functional of metric
$g$ on the manifold $\left(\mathcal{M},g\right)$), the MAG could
use more general forms of action. As some examples, the action Lagrangian
could be $\mathcal{L}\left(g,\boldsymbol{R}\right)$, a functional
of metric and the Riemann tensor $\boldsymbol{R}$ (and its contraction),
or $\mathcal{L}\left(g,\boldsymbol{R},\boldsymbol{T,Q}\right)$, where
the Lagrangian now is a functional of also torsion $T$ and non-metricity
$Q$ \cite{Iosifidis}. The second generalization is applied to the
connection $\nabla$ on $\left(\mathcal{M},g\right)$. Standard GR
adopts of a special case of connection known as Levi-Civita, where
the vector potential $\Gamma$ of $\nabla$ can always be written
as a function of metric $g$ on $\left(\mathcal{M},g\right)$. MAG,
on the other hand, uses a general affine connection $\omega$, which
in general, are independent of the metric \cite{Hehl,Smalley2,Smalley,Vitagliano,Iosifidis}.
The connection $\omega$ may contain non-metricity and torsional parts.
Because of the independence of the connection from the metric, the
action of MAG is also a functional on $\omega$, as well as a functional
of $g$. A special case of Metric-Affine Gravity with Einstein-Hilbert
action $\mathcal{S}\left[g,\omega\right]=\int\mathcal{R}\mathrm{vol}$
is known as Metric-Affine General Relativity (MAGR) or Generalized
Palatini gravity (however, these terminologies could be used differently,
as in \cite{tripathi}), while a more general families of such theories
with the $f(\mathcal{R})$ action, where $f(\mathcal{R})$ is the
polynomial functions of $\mathcal{R}$, is known as Metric-Affine-$f(\mathcal{R})$
Gravity \cite{Olmo,Sotiriou}. Both theories are considered as special
cases of MAG. The last generalization is on the existence of coupling
between matter and connection, which give rise to a new quantity known
as the \textit{hypermomentum}. This quantity vanishes in the standard
GR.

A detailed and complete explanation of MAG could be found in \cite{Vitagliano,Iosifidis},
while a detailed account of $f(\mathcal{R})$ and Metric-Affine $f(\mathcal{R})$
could be found in \cite{Oikonomou,Olmo,Sotiriou}. Some of the latest
progress on these theories include the mathematical and geometrical
analysis \cite{Iosifidis3,Iosifidis4}, ghost-free higher-order curvature
gravity \cite{Jimenez2,delhom,percacci2}, metric-affine cosmology
\cite{shimada,shimada2,mikura,damian1,damian2}, and the (3+1) formulation
of special cases of MAG \cite{Nathalie,Paschalidis,Shapiro,Olmo2,Iosifidis2,nuaing},
which become the main interest in this article.

As mentioned briefly in \cite{Eric}, the (3+1) formulation of General
Relativity was introduced by Darmois \cite{Darmois}, Lichnerowicz
\cite{Lichnerowicz1,Lichnerowicz2,Lichnerowicz3}, and Foures-Bruhat
\cite{Choquet1,Choquet2}, and then was popularized by Arnowitt, Deser,
and Misner in their famous article \cite{ADM}. In the (3+1) formulation,
the covariant fields on spacetime are split into their temporal and
spatial parts, to give rise to the dynamical variables on the 3D spatial
hypersurface. The (3+1) formulation for $f(\mathcal{R})$-gravity
for some special cases had been done in \cite{Nathalie,Paschalidis,Shapiro,Olmo2},
while some attempts on their quantization had been done in \cite{Ma1,Ma2,flavio}.
In our previous article \cite{nuaing}, we apply the (3+1) formulation
for the MAGR (complete with torsion and non-metricity), where the
focus is on the stress-energy-momentum equation (or the generalized
Einstein field equation), obtained by minimizing the action with respect
to the metric $g$. Here, as a companion article to \cite{nuaing},
the focus is on the hypermomentum equation, obtained from the minimization
of the action with respect to the connection $\omega$. The existence
of the hypermomentum $\mathcal{H}$ in the RHS of the hypermomentum
equation accomodates certain kinds of matter fields that give rise
to torsion and non-metricity. Another equation of interest in this
article is the traceless torsion constraint, which emerges from the
consequence of a consistent application of projective invariant constraint.
The origin of this constraint had been described in detail in \cite{Iosifidis,Iosifidis3,Iosifidis4}.
Another brief explanation could be seen in \cite{Sotiriou,Olmo,nuaing}.

The main result in this article is the (3+1) formulation of the hypermomentum
equation for the MAGR. Following our earlier work in \cite{nuaing},
we use the additional hypersurface variables with non-Riemannian origins
to write these (3+1) equations. The hypermomentum $\mathcal{H}$,
following \cite{Hehl2,Hehl3,Hehl4,Hehl5}, is decomposed into its
dilation, shear, and rotational (spin) parts. Moreover, we could show
that the hypermomentum could be constructed completely from  10 hypersurfaces
variables. In the end, by setting the hypermomentum to be zero, we
could retrieve the torsionless and metric compatibility in the (3+1)
framework, hence forcing the affine connection to be Levi-Civita.
This is possible with the help of the traceless torsion constraint
\cite{nuaing}.

The structure of this article is the following: In Section II, we
briefly review the geometrical setting and some definitions in \cite{nuaing},
these include the introduction of additional variables on the hypersurface.
Also, in this part of this article, we do the (3+1) decomposition
of the torsion $T$ and the non-metricity factor $\nabla_{X}g$, for
any $X\in T_{p}\mathcal{M}$.

Section III contains complete derivations on the equations of motion
of Metric-Affine $f\left(\mathcal{R}\right)$ gravity, i.e., the stress-energy-momentum
equation, obtained by minimizing $\mathcal{S}$ with respect to $g$,
and the hypermomentum equation, obtained by minimizing $\mathcal{S}$
with respect to $\omega$. The derivation of the hypermomentum equation
differs with the standard derivation in \cite{Sotiriou,Olmo}, however,
we show that both ways are equivalent. The traceless torsion constraint
is also introduced in this section. At the end of this section, we
consider a special case of Metric-Affine $f\left(\mathcal{R}\right)$-gravity,
where $f\left(\mathcal{R}\right)=\mathcal{R}$, i.e., the Metric-Affine
General Relativity (MAGR) or Generalized Palatini gravity. From this
stage up to the rest of this article, we will only conider this case.

In Section IV, we perform the (3+1) decomposition to the hypermomentum
equation. The first step is to split the hypermomentum $\mathcal{H}$
in the RHS of the hypermomentum equation into its temporal and spatial
parts. The split of the hypermomentum could be simplified further
by decomposing $\mathcal{H}$ into the dilation, shear, and rotational
(spin) parts. With these, the hypermomentum could be constructed completely
from 10 hypersurfaces variables. The second step is to split the LHS
of the hypermomentum equation using some (3+1) relations we derived
in Section II. The results are (3+1) hypermomentum equations, which
consist of 1 scalar equation, 3 vector equations, 3 matrix equations,
and 1 tensor equation of order $\tbinom{2}{1}$. Moreover, we perform
the (3+1) decomposition of the traceless torsion constraint, resulting
in 1 scalar and 1 vector equation; collecting these results altogether,
we obtain 10 hypersurface equations, which are the main result in
this article.

In Section V, we consider some special cases of MAGR. For the first
case, we set the hypermomentum $\mathcal{H}=0$, and show that this
gives  the metric compatibility and torsionless condition of the Levi-Civita
connection. For the second and third case, we apply the metric compatibility
and torsionless condition separately to the (3+1) hypermomentum equation;
for these two cases, the hypermomentum, as well as the dynamical equations,
are restricted by constraints. 

Finally, we discussed some subtleties on the results and conclude
our works in Section VI.

\section{The Geometrical Setting and Definitions}

The geometrical setting is similar with the one we used in our previous
companion article \cite{nuaing}. Let $\mathcal{M}$ be a $4$-dimensional,
globally hyperbolic, Lorentzian manifold, equipped with a Lorentzian
metric $g$ and an affine connection $\nabla$ that are independent
from $g$. Let $\Sigma$ be the 3-dimensional (Riemannian) hypersurface
on $\mathcal{M},$ and $\hat{n}_{p}$ be the unit vector normal to
$\Sigma$ at each point $p\in\Sigma$ satisfying $g_{p}\left(\hat{n}_{p},\hat{n}_{p}\right)=-1$.
The metric $g$ on $\mathcal{M}$ induces a Riemannian metric $\,^{3}q$
on the hypersurface $\Sigma$:
\begin{equation}
\,^{3}q=g+\hat{n}^{*}\otimes\hat{n}^{*},\label{eq:1a}
\end{equation}
with $\hat{n}^{*}\in T_{p}^{*}\mathcal{M}$ is the covariant vector
to $\hat{n},$ satisfying $\hat{n}^{*}=g\left(\hat{n},\cdot\right)=g\left(\hat{n}\right)$
(the label $p$ is omitted for simplicity).

\subsubsection*{The Adapted Coordinate, Lapse Function, and Shift Vector}

Let $x^{\mu}=\left\{ x^{0},x^{i}\right\} $ be a local coordinate
on $\mathcal{M}$, with $x^{0}$ and $x^{i}$ are, respectively, the
temporal and spatial part of $x^{\mu}$. The corresponding coordinate
vector basis on $T_{p}\mathcal{M}$ is $\partial_{\mu}=\left\{ \partial_{0},\partial_{i}\right\} $.
Any vector $V\in T_{p}\mathcal{M}$ could be decomposed as follows:
\[
V=\underset{V^{\mu}}{\underbrace{\left\langle dx^{\mu},V\right\rangle }}\partial_{\mu},
\]
with $V^{\mu}$ is the component of $V$. $dx^{\mu}$ is the coordinate
covector basis on $T_{p}^{*}\mathcal{M}$ satisfying $\left\langle dx^{\mu},\partial_{\nu}\right\rangle =\partial_{\nu}\left(dx^{\mu}\right)=\delta_{\nu}^{\mu}$
(or $dx^{\mu}=g^{\mu\nu}g\left(\partial_{\nu}\right)$). Notice that
$g_{\mu\nu}=g\left(\partial_{\mu},\partial_{\nu}\right)=\left\langle g\left(\partial_{\mu}\right),\partial_{\nu}\right\rangle $.

Let us consider another set of vector basis which spans $T_{p}\mathcal{M}$,
namely $\left(\hat{n},\partial_{i}\right)$, where $\partial_{i}$
is simultaneously the coordinate vector basis of the hypersurface
$\Sigma$. As a consequence, $g\left(\hat{n},\partial_{i}\right)=n_{i}=0$.
$\left(\hat{n},\partial_{i}\right)$ is known as the adapted vector
basis to the hypersurface $\Sigma$. In this basis, the temporal component
of $\partial_{\mu},$ $\partial_{0},$ is decomposed into:
\begin{equation}
\partial_{0}=\underset{N}{\underbrace{-\left\langle \hat{n}^{*},\partial_{0}\right\rangle }\hat{n}}+\underset{N^{i}}{\underbrace{\left\langle \,^{3}dx^{i},\partial_{0}\right\rangle \partial_{i}}},\label{eq:lapseshift}
\end{equation}
where $\,^{3}dx^{i}=\,^{3}q^{ij}\,^{3}q\left(\partial_{j}\right),$
is the covector basis which spans $T_{p}^{*}\Sigma$. $N$ and $\boldsymbol{N}=N^{i}\partial_{i}$
are, respectively, the lapse function and the shift vector of $\partial_{0}$.
By solving $\left\langle dx^{\mu},\partial_{\nu}\right\rangle =\delta_{\nu}^{\mu}$,
one could obtain that:
\begin{align}
dx^{0}= & -\hat{n}^{*}N^{-1},\label{eq:lapshift2}\\
dx^{i}= & \hat{n}^{*}N^{i}N^{-1}+\,^{3}dx^{i}.\label{eq:lapshift3}
\end{align}
Using (\ref{eq:lapseshift}), (\ref{eq:lapshift2}), and (\ref{eq:lapshift3}),
the components of metric $g$ and its inverse $g^{-1}=g^{*}$ could
be written in terms of lapse and shift as follows:
\begin{center}
\begin{tabular}{lcl}
$g\left(\partial_{0},\partial_{0}\right)=g_{00},=N^{i}N_{i}-N^{2},$ & $\qquad$ & $g^{*}\left(dx^{0},dx^{0}\right)=g^{00}=-N^{-2},$\tabularnewline
$g\left(\partial_{0},\partial_{i}\right)=g_{0i},=N_{i},$ &  & $g^{*}\left(dx^{0},dx^{i}\right)=g^{0i}=N^{i}N^{-2},$\tabularnewline
$g\left(\partial_{i},\partial_{j}\right)=g_{ij}=\,^{3}q_{ij},$ &  & $g^{*}\left(dx^{i},dx^{j}\right)=g^{ij}=\,^{3}q^{ij}-\left(N^{i}N^{j}\right)N^{-2},$\tabularnewline
\end{tabular}
\par\end{center}

where $\,^{3}q^{ij}$ satisfies $\,^{3}q^{ij}=g^{*}\left(\,^{3}dx^{i},\,^{3}dx^{j}\right)$.
The following equations will be useful for the later derivations in
this article \cite{nuaing}:
\begin{center}
\begin{tabular}{lcl}
$\hat{n}=n^{0}\partial_{0}+n^{i}\partial_{i}=N^{-1}\left(\partial_{0}-N^{i}\partial_{i}\right),$ & $\qquad$ & $\boldsymbol{N}=N^{0}\partial_{0}+N^{i}\partial_{i}=-Nn^{i}\partial_{i},$\tabularnewline
$\hat{n}^{*}=n_{0}dx^{0}+n_{i}dx^{i}=-Ndx^{0},$ &  & $\boldsymbol{N}^{*}=N_{0}dx^{0}+N_{i}dx^{i}=N^{i}N_{i}dx^{0}+\,^{3}q_{ij}N^{j}dx^{i}.$\tabularnewline
\end{tabular}
\par\end{center}

\subsubsection*{The Additional Variables}

Following \cite{nuaing}, all the 3+1 quantities will be written in
terms of the additional variables motivated by the concept of geometrodynamics
introduced in \cite{Wheeler}, defined as follows:
\begin{align}
\nabla_{\hat{n}}\hat{n} & =\Theta\left(\hat{n}\right)\hat{n}+\alpha^{i}\partial_{i},\label{eq:a}\\
\nabla_{i}\hat{n} & =\Theta_{i}\hat{n}+\mathcal{K}_{i}^{\:\:j}\partial_{j},\label{eq:b}\\
\nabla_{i}\partial_{j} & =K_{ij}\hat{n}+\,^{3}\omega_{i\:\,\:j}^{\:\:k}\partial_{k},\label{eq:c}\\
\nabla_{\hat{n}}\partial_{i} & =\Delta_{i}\hat{n}+\Delta_{\;\:i}^{j}\partial_{j}.\label{eq:d}
\end{align}
$\left(\Theta\left(\hat{n}\right),^{3}\alpha=\alpha^{i}\right)$ are
the temporal and spatial components of the 4-acceleration, $K$ and
$\mathcal{K}$ are extrinsic curvatures of the first and second kind,
$\Theta_{i}$ is the angle between the 3-acceleration $\alpha_{i}$
and the normal $\hat{n}$, $\,^{3}\omega$ is the 3-connection. $\Delta_{i}$
and $\Delta_{\;\:i}^{j}$ are the temporal and spatial components
of the evolution generator of $\partial_{i}$. The quantities, in
general, have non-Riemannian origins, for a detailed explanation,
consult \cite{nuaing}.

\subsubsection*{(3+1)-Decomposition of the Torsion Tensor}

Due to the different conventions used in the literature, we need to
fix the notations used in this article. The connection is written
as $\nabla_{\mu}\partial_{\beta}=\omega_{\mu\,\:\,\beta}^{\,\:\,\alpha}\partial_{\alpha}$,
the torsion tensor as $\boldsymbol{T}\left(\partial_{\mu},\partial_{\beta}\right)=T_{\mu\,\:\,\beta}^{\:\:\alpha}\partial_{\alpha}$,
and the generalized Riemann tensor as $\boldsymbol{R}\left(\partial_{\mu},\partial_{\nu}\right)\partial_{\beta}=R_{\mu\nu\,\:\,\,\beta}^{\:\:\:\:\,\,\alpha}\partial_{\alpha},$
which are different from the ones used in, for examples \cite{Olmo,Sotiriou}.

The torsion $T\left(\partial_{\mu},\partial_{\lambda}\right)=T_{\mu\;\;\;\lambda}^{\:\:\alpha}\partial_{\alpha}\in T_{p}\mathcal{M}$
is defined as:
\begin{equation}
T\left(X,Y\right)=\nabla_{X}Y-\nabla_{Y}X-\left[X,Y\right],\qquad X,Y\in T_{p}\mathcal{M}.\label{eq:definitiontorsion}
\end{equation}
The torsion is decomposed into temporal and spatial parts, namely
the quantity $T\left(\hat{n},\hat{n}\right),$ $T\left(\hat{n},\partial_{i}\right)=-T\left(\partial_{i},\hat{n}\right),$
and $T\left(\partial_{i},\partial_{j}\right)$. From the definition
(\ref{eq:definitiontorsion}), it is clear that:
\begin{equation}
T\left(\hat{n},\hat{n}\right)=0.\label{eq:T00}
\end{equation}
 Using the additional variables, one could obtain that:
\begin{equation}
T\left(\hat{n},\partial_{i}\right)=-T\left(\partial_{i},\hat{n}\right)=\left(\Delta_{i}-\Theta_{i}-\left\langle \hat{n}^{*},\partial_{i}\hat{n}\right\rangle \right)\hat{n}+\left(\Delta_{\;\:i}^{j}-\mathcal{K}_{i}^{\:\:j}+\left\langle \,^{3}dx^{j},\partial_{i}\hat{n}\right\rangle \right)\partial_{j},\label{eq:T0i}
\end{equation}
and:
\begin{equation}
T\left(\partial_{i},\partial_{j}\right)=\,^{3}T\left(\partial_{i},\partial_{j}\right)+\left(K\left(\partial_{i},\partial_{j}\right)-K\left(\partial_{j},\partial_{i}\right)\right)\hat{n}.\label{eq:Tij}
\end{equation}
(\ref{eq:T00}), (\ref{eq:T0i}), and (\ref{eq:Tij}) are the (3+1)
decomposition of $T\left(\partial_{\mu},\partial_{\lambda}\right)$.

Let us define the component of torsion $T_{\mu}^{\:\:\,\alpha\beta}$
as follows:
\begin{align}
T_{\mu}^{\:\:\,\alpha\beta}\partial_{\alpha} & =g^{\lambda\beta}T_{\mu\;\;\;\lambda}^{\:\,\:\alpha}\partial_{\alpha}=g^{\lambda0}T\left(\partial_{\mu},\partial_{0}\right)\partial_{\alpha}+g^{\lambda i}T\left(\partial_{\mu},\partial_{i}\right)\partial_{\alpha},\label{eq:raised}
\end{align}
where the last index of $T_{\mu\;\;\;\lambda}^{\:\:\,\alpha}$ is
raised by the inverse metric $g^{*}.$ An important quantity for the
derivation in the next sections are some contractions of (\ref{eq:raised})
as follows, that are obtained with the help of (\ref{eq:raised}):
\begin{align}
n^{\mu}n_{\beta}T_{\mu}^{\:\:\alpha\beta}\partial_{\alpha} & =n^{\mu}n^{\beta}T_{\mu\;\;\;\beta}^{\:\:\alpha}\partial_{\alpha}=T\left(\hat{n},\hat{n}\right),\label{eq:p}\\
n^{\mu}T_{\mu}^{\:\:\alpha i}\partial_{\alpha} & =n^{\mu}g^{i\lambda}T_{\mu\;\;\;\lambda}^{\:\:\alpha}\partial_{\alpha}=g^{i\lambda}T\left(\hat{n},\partial_{\lambda}\right)\partial_{\alpha}=N^{i}N^{-1}T\left(\hat{n},\hat{n}\right)+\,^{3}q^{ij}T\left(\hat{n},\partial_{j}\right),\label{eq:q}\\
n_{\beta}T_{i}^{\:\:\alpha\beta}\partial_{\alpha} & =n^{\beta}T_{i\;\;\;\beta}^{\:\:\alpha}\partial_{\alpha}=T\left(\partial_{i},\hat{n}\right),\label{eq:r}\\
T_{i}^{\:\:\alpha j}\partial_{\alpha} & =g^{j\beta}T_{i\;\;\;\lambda}^{\:\:\alpha}\partial_{\alpha}=N^{j}N^{-1}T\left(\partial_{i},\hat{n}\right)+\,^{3}q^{jk}T\left(\partial_{i},\partial_{k}\right).\label{eq:s}
\end{align}
Inserting (\ref{eq:T00}), (\ref{eq:T0i}), and (\ref{eq:Tij}) to
these 4 equations gives:
\begin{align*}
n^{\mu}n_{\beta}T_{\mu}^{\:\:\alpha\beta}\partial_{\alpha}= & 0,\\
n^{\mu}T_{\mu}^{\:\:\alpha i}\partial_{\alpha}= & \,^{3}q^{ij}\left(\Delta_{j}-\Theta_{j}-\left\langle \hat{n}^{*},\partial_{j}\hat{n}\right\rangle \right)\hat{n}+\,^{3}q^{ij}\left(\Delta_{\;\:j}^{k}-\mathcal{K}_{j}^{\:\:k}+\left\langle \,^{3}dx^{k},\partial_{j}\hat{n}\right\rangle \right)\partial_{k},\\
n_{\beta}T_{i}^{\:\:\alpha\beta}\partial_{\alpha}= & -\left(\Delta_{i}-\Theta_{i}-\left\langle \hat{n}^{*},\partial_{i}\hat{n}\right\rangle \right)\hat{n}+\left(\Delta_{\;\:i}^{j}-\mathcal{K}_{i}^{\:\:j}+\left\langle \,^{3}dx^{j},\partial_{i}\hat{n}\right\rangle \right)\partial_{j},\\
T_{i}^{\:\:\alpha j}\partial_{\alpha}= & \left(-N^{j}N^{-1}\left(\Delta_{i}-\Theta_{i}-\left\langle \hat{n}^{*},\partial_{i}\hat{n}\right\rangle \right)+\,^{3}q^{jk}\left(K_{ik}-K_{ki}\right)\right)\hat{n}\\
 & \;-N^{j}N^{-1}\left(\Delta_{\;\:i}^{k}-\mathcal{K}_{i}^{\:\:k}+\left\langle \,^{3}dx^{k},\partial_{i}\hat{n}\right\rangle \right)\partial_{k}+\,^{3}q^{jk}\,^{3}T\left(\partial_{i},\partial_{k}\right).
\end{align*}

\subsubsection*{Decomposition of the Non-Metricity Factor}

One could decompose the non-metricity factor $\nabla_{X}g$ into its
temporal and spatial parts, for every $X\in T_{p}\mathcal{M};$ however,
considering the form of equations (\ref{eq:a})-(\ref{eq:d}), it
is more convenient to obtain the (3+1) decomposition of the quantity
$\nabla_{X}g^{*}$. With the decomposition of metric in (\ref{eq:d}),
one could write:
\[
\nabla_{X}g^{*}=\nabla_{X}\left(\,^{3}q^{*}-\hat{n}\otimes\hat{n}\right).
\]
Using (\ref{eq:a})-(\ref{eq:d}), one could obtain:
\begin{align}
\nabla_{\hat{n}}g^{*} & =-2\Theta\left(\hat{n}\right)\hat{n}\otimes\hat{n}+\left(\Delta^{i}-\alpha^{i}\right)\left(\hat{n}\otimes\partial_{i}+\partial_{i}\otimes\hat{n}\right)+\,^{3}\nabla_{\hat{n}}\,^{3}q^{*},\label{eq:q1}\\
\nabla_{i}g^{*} & =-2\Theta_{i}\hat{n}\otimes\hat{n}+\left(K_{i}^{\;j}-\mathcal{K}_{i}^{\:\:j}\right)\left(\hat{n}\otimes\partial_{j}+\partial_{j}\otimes\hat{n}\right)+\,^{3}\nabla_{i}\,^{3}q^{*},\label{eq:q2}
\end{align}
where:
\begin{align}
\,^{3}\nabla_{\hat{n}}\,^{3}q^{*}=\left(\,^{3}\nabla_{\hat{n}}\,^{3}q^{ij}\right)\partial_{i}\otimes\partial_{j}= & \left(\hat{n}\left[\,^{3}q^{ij}\right]+\Delta_{\;\:k}^{i}\,^{3}q^{kj}+\Delta_{\;\:k}^{j}\,^{3}q^{ki}\right)\partial_{i}\otimes\partial_{j},\label{eq:q3}\\
\,^{3}\nabla_{i}\,^{3}q^{*}=\left(\,^{3}\nabla_{i}\,^{3}q^{jk}\right)\partial_{i}\otimes\partial_{k}= & \left(\partial_{i}\,^{3}q^{jk}+\omega_{i}^{\;jk}+\omega_{i}^{\;kj}\right)\partial_{i}\otimes\partial_{k}.\label{eq:q4}
\end{align}

The following are the list of useful relations for the derivation
of the (3+1) hypermomentum equation. In indices, $\nabla_{X}g^{*}$
could be written as: 
\[
\nabla_{X}g^{*}=X^{\mu}\left(\nabla_{\mu}g^{\alpha\beta}\right)\partial_{\alpha}\otimes\partial_{\beta},
\]
where $\nabla_{\mu}g^{\alpha\beta}=\partial_{\mu}g^{\alpha\beta}+\omega_{\mu}^{\;\:\alpha\beta}+\omega_{\mu}^{\;\:\beta\alpha}.$
Hence, for $Y,Z\in T_{p}^{*}\mathcal{M}$, one has:
\begin{align}
\left(\nabla_{X}g^{*}\right)\left(dx^{\mu},dx^{\nu}\right) & =X^{\sigma}\nabla_{\sigma}g^{\mu\nu},\nonumber \\
\left(\nabla_{X}g^{*}\right)\left(Y^{*},Z^{*}\right) & =X^{\mu}Y_{\alpha}Z_{\beta}\nabla_{\mu}g^{\alpha\beta},\label{eq:h}\\
\left(\nabla_{X}g^{*}\right)\left(Y^{*},\cdot\right)= & \left(\nabla_{X}g^{*}\right)\left(\cdot,Y^{*}\right)=\left(X^{\mu}Y_{\alpha}\nabla_{\mu}g^{\alpha\beta}\right)\partial_{\beta}.\label{eq:i}
\end{align}
Moreover, one has:
\begin{align}
\nabla_{\mu}g^{\mu\beta} & =-\left(\nabla_{n}g^{*}\right)\left(\hat{n}^{*},dx^{\beta}\right)-N^{-1}N^{i}\left(\nabla_{i}g^{*}\right)\left(\hat{n}^{*},dx^{\beta}\right)+\left(\nabla_{i}g^{*}\right)\left(dx^{i},dx^{\beta}\right),\label{eq:f}\\
g^{\alpha\beta}\nabla_{\mu}g_{\alpha\beta} & =-g_{\alpha\beta}\nabla_{\mu}g^{\alpha\beta}=-\left(N^{-2}N^{i}N_{i}-1\right)\left(\nabla_{\mu}g^{*}\right)\left(\hat{n}^{*},\hat{n}^{*}\right)+2N^{-1}N_{i}\left(\nabla_{\mu}g^{*}\right)\left(\hat{n}^{*},dx^{i}\right)-\,^{3}q_{ij}\left(\nabla_{\mu}g^{*}\right)\left(dx^{i},dx^{j}\right).\label{eq:g}
\end{align}
The last quantity is known as the Weyl vector $Q_{\mu}=-g^{\alpha\beta}\nabla_{\mu}g_{\alpha\beta}$.
One needs to be careful that (\ref{eq:f})-(\ref{eq:g}) are written
in $dx^{i}$ instead of $^{3}dx^{i}.$ A complete (3+1) decomposition
of (\ref{eq:f}) and (\ref{eq:g}) could be obtained by inserting
(\ref{eq:lapshift3}), but this is not necessary for the moment.

\section{The Metric-Affine $f\left(R\right)$-gravity}

Let us review briefly the theory of Metric-Affine $f\left(R\right)$-Gravity;
for some detailed and complete derivation, one could consult \cite{Iosifidis,Olmo,Sotiriou}.
The corresponding action is defined as follows:
\begin{equation}
S\left[g,\omega,\chi,\zeta\right]=S_{\mathrm{GR}}\left[g,\omega\right]+S_{\mathrm{matter}}\left[g,\omega\right]+S_{\mathrm{LM}}\left[\chi,g,\omega\right]+S_{\mathrm{sym}}\left[\zeta,g\right].\label{eq:MAfR}
\end{equation}
The first term $S_{\mathrm{GR}}\left[g,\omega\right]$ is the gravitational
term:
\[
S_{\mathrm{GR}}\left[g,\omega\right]=\intop_{\mathcal{M}}f\left(\mathcal{R}\left[\omega\right]\right)\mathrm{vol},
\]
$f\left(\mathcal{R}\right)$ is the polynomials of Ricci scalar:
\[
f\left(\mathcal{R}\right)=..+\frac{c_{2}}{\mathcal{R}^{2}}+\frac{c_{1}}{\mathcal{R}}-2\Lambda+\mathcal{R}+\frac{\mathcal{R}^{2}}{k_{2}}+\frac{\mathcal{R}^{3}}{k_{3}}+...
\]
with:
\begin{align}
\boldsymbol{R}\left(\partial_{\mu},\partial_{\mu}\right)\partial_{\beta} & =\nabla_{\mu}\nabla_{\nu}\partial_{\beta}-\nabla_{\nu}\nabla_{\mu}\partial_{\beta}=\underset{R_{\mu\nu\,\:\,\beta}^{\:\:\:\:\,\,\alpha}}{\underbrace{\left(\partial_{\mu}\omega_{\nu\,\:\beta}^{\:\:\,\alpha}-\partial_{\nu}\omega_{\mu\,\:\beta}^{\:\:\,\alpha}+\omega_{\mu\,\:\sigma}^{\:\,\:\alpha}\omega_{\nu\,\:\beta}^{\:\:\,\sigma}-\omega_{\nu\,\:\sigma}^{\:\,\:\alpha}\omega_{\mu\,\:\beta}^{\:\:\,\sigma}\right)}\partial_{\alpha}.}\label{eq:riemann}\\
\mathbf{Ric} & =\left\langle dx^{\mu},\boldsymbol{R}\left(\partial_{\mu},\partial_{\nu}\right)\partial_{\beta}\right\rangle dx^{\nu}\otimes dx^{\beta}=\underset{R_{\nu\beta}}{\underbrace{\delta_{\alpha}^{\mu}R_{\mu\nu\,\:\,\beta}^{\:\:\:\:\,\,\alpha}}}dx^{\nu}\otimes dx^{\beta},\label{eq:ricci}\\
\mathcal{R} & =\mathrm{tr}\,\mathbf{Ric}=g^{\nu\beta}R_{\nu\beta}.\label{eq:scalar}
\end{align}
$\boldsymbol{R}$, $\mathbf{Ric},$ and $\mathcal{R}$ are the generalized
Riemann curvature tensor, Ricci tensor, and (Ricci) scalar curvature.
The action $S$ is a functional over the metric $g$, connection $\omega$,
an a Lagrange multiplier $\chi$. The connection $\omega$ is a general
affine connection and hence is independent from the metric $g$ \cite{Iosifidis,Olmo,Sotiriou}.
The second term on the RHS of (\ref{eq:MAfR}), $S_{\mathrm{matter}}\left[g,\omega\right]$,
is the Lagrangian of the matter density, which is a functional of
$g$ and $\omega$, while $S_{LM}$ is the Lagrange multiplier term,
defined as follows:
\[
S_{LM}=\intop_{\mathcal{M}}\chi^{\mu}T_{\mu}\textrm{vol },
\]
$\chi^{\mu}$ is the Lagrange multiplier corresponding to the traceless
torsion constraint $T_{\mu}=T_{\alpha\;\:\mu}^{\;\,\alpha}=0$ \cite{Iosifidis,Olmo,Sotiriou}.
The traceless torsion constraint is introduced to guarantee the consistency
of the equations of motion under the projective invariance transformation,
see \cite{Iosifidis,Olmo,Sotiriou}. As discovered in \cite{Iosifidis3,Iosifidis4},
the constraint $T_{\mu}=0$ is not the only possible condition to
guarantee the consistency, one could equally choose the Weyl tensor
$Q_{\mu}=0$, or moreover, the linear combination between these two.
In this article, we choose the constraint $T_{\mu}=0$, which is frequently
used in the literature. The treatment where $Q_{\mu}=0$ could be
found elsewhere.

The last term in (\ref{eq:MAfR}), $S_{\mathrm{sym}}$, is another
Lagrange multiplier term related to another 'constraint' of the system,
namely, the symmetricity of the metric tensor. One should note that
for a general affine connection, the Ricci tensor (\ref{eq:ricci})
is not symmetric by the interchange of the $\left(\nu\beta\right)$
indices, in other words, the Ricci tensor has an antisymmetric part
$R_{\nu\beta}=R_{\left(\nu\beta\right)}+R_{\left[\nu\beta\right]}$
(here respectively, we use $\left(.,.\right)$ and $\left[.,.\right]$
to label symmetricity and antisymmetricity). However, this antisymmetric
part does not contribute to the Ricci scalar $\mathcal{R}$ since
the contraction of $R_{\left[\nu\beta\right]}$ with $g^{\nu\beta}$
is always zero. This is due to the fact that the metric is always
symmetric, namely $g^{\left[\nu\beta\right]}=0$. To obtain consistent
equations of motion, one needs as well to consider this constraint
as follows:
\[
S_{\mathrm{sym}}=\intop_{\mathcal{M}}\zeta_{\alpha\beta}g^{\left[\alpha\beta\right]}\textrm{vol },
\]
with $\zeta$ is the Lagrange multiplier corresponding to the constraint
$g^{\left[\alpha\beta\right]}=0$.

\subsubsection*{The Stress-Energy-Momentum Equation}

In this subsection, we will briefly review the derivation of the stress-energy-momentum
tensor; other variations in deriving the results could be found on
the literatures. The variation of a functional $G\left[x\right]$
with respect to a variable $x$ is defined as:
\begin{equation}
\delta_{x}G=\left.\frac{d}{ds}G\left[x+s\delta x\right]\right|_{s=0}.\label{eq:def}
\end{equation}
With this relation in mind, we will derive all the equations of motions
of the action (\ref{eq:MAfR}). The variation of the action $S$ with
respect to $g$ assuming the variation of $\omega$, $\chi$, and
$\zeta$ are zero, gives:
\begin{align}
\delta_{g}S & =\intop_{\mathcal{M}}\left(\delta_{g}f\left(\mathcal{R}\left[\omega\right]\right)\right)\mathrm{vol}+f\left(\mathcal{R}\left[\omega\right]\right)\delta_{g}\mathrm{vol}+\delta_{g}S_{\mathrm{matter}}\left[g,\omega\right]+\delta_{g}S_{LM}+\delta_{g}S_{\mathrm{sym}}.\label{eq:dS}
\end{align}
The volume form ($\mathrm{vol}$) for the Lorentzian metric is $\mathrm{vol}=\sqrt{\left|-\det g\right|}d^{n}x,$
which is only a function of $g$, and by a standard calculation, it
gives:
\begin{equation}
\delta_{g}\mathrm{vol}=-\frac{1}{2}g_{\alpha\beta}\left(\delta g^{\alpha\beta}\right)\mathrm{vol}.\label{eq:satu}
\end{equation}
The variation of $f\left(\mathcal{R}\right)$, with (\ref{eq:def})
and the chain rule, gives:
\[
\delta_{g}f\left(\mathcal{R}\right)=f'\left(\mathcal{R}\right)\delta_{g}\mathcal{R},
\]
with $f'\left(\mathcal{R}\right)=\frac{df}{d\mathcal{R}}$. 

Next, the variation of the Ricci scalar could be calculated as follows:
\[
\delta_{g}\mathcal{R}=\delta_{g}\left(g^{\alpha\beta}R_{\alpha\beta}\right)=\left(\delta g^{\alpha\beta}\right)R_{\alpha\beta}+g^{\alpha\beta}\underset{0}{\underbrace{\delta_{g}R_{\alpha\beta}}},
\]
where $\delta_{g}R_{\alpha\beta}=0,$ since for a general affine connection,
the Ricci tensor $R_{\alpha\beta}$ is only a function of $\omega.$
Therefore, by the variation of $\delta g$, we have:
\begin{equation}
\delta_{g}f\left(\mathcal{R}\right)=f'\left(\mathcal{R}\right)R_{\alpha\beta}\delta g^{\alpha\beta}.\label{eq:dua}
\end{equation}
Moreover, the variation of the matter Lagrangian with respect to the
metric is defined as:
\begin{equation}
\delta_{g}S_{\mathrm{matter}}\left[g,\omega\right]:=-\intop_{\mathcal{M}}\kappa\mathcal{T}_{\alpha\beta}\delta g^{\alpha\beta}\mathrm{vol},\label{eq:tiga}
\end{equation}
with $\mathcal{T}$ (not to be confused with the torsion $T$) is
the standard stress-energy-momentum tensor, symmetric in the $\left(\alpha,\beta\right)$
indices \cite{Sotiriou}.

The variation of $S_{LM}$ with respect to $g$ is:
\[
\delta_{g}S_{LM}=\intop_{\mathcal{M}}\left(\delta_{g}\chi^{\mu}\right)T_{\alpha\;\:\mu}^{\;\,\alpha}\mathrm{vol}+\intop_{\mathcal{M}}\chi^{\mu}\left(\delta_{g}T_{\alpha\;\:\mu}^{\;\,\alpha}\right)\mathrm{vol}+\intop_{\mathcal{M}}\chi^{\mu}T_{\alpha\;\:\mu}^{\;\,\alpha}\left(\delta_{g}\mathrm{vol}\right).
\]
Both the first and second terms are zero because the Lagrange multiplier
$\chi^{\mu}$ and the trace of torsion $T_{\mu}$ are independent
from the metric $g$. Therefore, using (\ref{eq:satu}):
\begin{equation}
\delta_{g}S_{LM}=-\intop_{\mathcal{M}}\frac{1}{2}\chi^{\mu}T_{\nu\;\:\mu}^{\;\,\,\nu}g_{\alpha\beta}\left(\delta g^{\alpha\beta}\right)\mathrm{vol}.\label{eq:empat}
\end{equation}

Finally, the variation of $S_{\mathrm{sym}}$ with respect to $g$
is:
\[
\delta_{g}S_{\mathrm{sym}}=\intop_{\mathcal{M}}\underset{0}{\underbrace{\left(\delta_{g}\zeta_{\alpha\beta}\right)}}g^{\left[\alpha\beta\right]}\textrm{vol }+\intop_{\mathcal{M}}\zeta_{\alpha\beta}\left(\delta_{g}g^{\left[\alpha\beta\right]}\right)\textrm{vol }+\intop_{\mathcal{M}}\zeta_{\alpha\beta}g^{\left[\alpha\beta\right]}\left(\delta_{g}\textrm{vol }\right).
\]
The first term is zero because the Lagrange multiplier $\zeta_{\alpha\beta}$
is independent from the metric $g$. The second term is:
\[
\delta_{g}g^{\left[\alpha\beta\right]}=\frac{1}{2}\delta_{g}\left(g^{\alpha\beta}-g^{\beta\alpha}\right)=\frac{1}{2}\left(\delta_{\mu}^{\alpha}\delta_{\nu}^{\beta}-\delta_{\mu}^{\beta}\delta_{\nu}^{\alpha}\right)\delta g^{\mu\nu}.
\]
Therefore, using (\ref{eq:satu}):
\begin{equation}
\delta_{g}S_{\mathrm{sym}}=\frac{1}{2}\intop_{\mathcal{M}}\left(\zeta_{\alpha\beta}-\zeta_{\beta\alpha}-\zeta_{\mu\nu}g^{\left[\mu\nu\right]}g_{\alpha\beta}\right)\left(\delta g^{\alpha\beta}\right)\textrm{vol }.\label{eq:lima}
\end{equation}
Inserting (\ref{eq:satu}), (\ref{eq:dua}), (\ref{eq:tiga}), (\ref{eq:empat}),
and (\ref{eq:lima}) to (\ref{eq:dS}) gives:
\[
\delta_{g}S=\intop_{\mathcal{M}}\left(f'\left(\mathcal{R}\right)R_{\alpha\beta}-\frac{1}{2}f\left(\mathcal{R}\right)g_{\alpha\beta}-\kappa\mathcal{T}_{\alpha\beta}-\frac{1}{2}\left(\chi^{\mu}T_{\nu\;\:\mu}^{\;\,\,\nu}+\zeta_{\mu\nu}g^{\left[\mu\nu\right]}\right)g_{\alpha\beta}+\zeta_{\left[\alpha\beta\right]}\right)\delta g^{\alpha\beta}\mathrm{vol}.
\]

The variation $\delta_{g}S$ is zero for any variation of $\delta g^{\alpha\beta}$
if the following Euler-Lagrange equation is satisfied:
\begin{equation}
\underset{G_{\alpha\beta}}{\underbrace{f'\left(\mathcal{R}\right)R_{\alpha\beta}-\frac{1}{2}f\left(\mathcal{R}\right)g_{\alpha\beta}-\kappa\mathcal{T}_{\alpha\beta}-\frac{1}{2}\left(\chi^{\mu}T_{\nu\;\:\mu}^{\;\,\,\nu}+\zeta_{\mu\nu}g^{\left[\mu\nu\right]}\right)g_{\alpha\beta}+\zeta_{\left[\alpha\beta\right]}}}=\kappa\mathcal{T}_{\alpha\beta}.\label{eq:EL1}
\end{equation}
$\boldsymbol{G}$ is known as the generalized Einstein tensor, and
(\ref{eq:EL1}) is the stress-energy-momentum equation of Metric-Affine
$f\left(\mathcal{R}\right)$-Gravity.

\subsubsection*{The Hypermomentum Equation}

The second equation of motion could be obtained from the variation
of $S$ with respect to the affine connection $\omega$, $\delta_{\omega}S$,
now assuming the variation of $g$, $\chi$, and $\zeta$ are zero:
\begin{align}
\delta_{\omega}S & =\intop_{\mathcal{M}}\left(\delta_{\omega}f\left(\mathcal{R}\left[\omega\right]\right)\right)\mathrm{vol}+f\left(\mathcal{R}\left[\omega\right]\right)\underset{0}{\underbrace{\delta_{\omega}\mathrm{vol}}}+\delta_{\omega}S_{\mathrm{matter}}\left[g,\omega\right]+\delta_{\omega}S_{LM}+\delta_{\omega}S_{\mathrm{sym}}.\label{eq:dS-1}
\end{align}
We will carry the derivation of the hypermomentum equation differently
with the standard derivation in \cite{Olmo,Sotiriou}, in order to
have a more simpler form for the (3+1) decomposition. As we had mentioned
in the previous derivation, $\delta_{\omega}\mathrm{vol}$ is zero
since it is only a function of $g$. Similar with (\ref{eq:dua}),
we have:
\begin{equation}
\delta_{\omega}f\left(\mathcal{R}\right)=f'\left(\mathcal{R}\right)\delta_{\omega}\mathcal{R}.\label{eq:x}
\end{equation}
The quantity $\delta_{\omega}\mathcal{R}$ could be derived as follows.
With definition (\ref{eq:riemann}), one could show that the variation
of Ricci tensor (\ref{eq:ricci}) with respect to $\omega$ is:
\begin{equation}
\delta_{\omega}R_{\alpha\beta}=\delta_{\omega}R_{\mu\alpha\,\:\,\beta}^{\:\:\:\:\:\,\mu}=\nabla_{\mu}\delta\omega_{\alpha\,\:\beta}^{\:\:\,\mu}-\nabla_{\alpha}\delta\omega_{\mu\,\:\beta}^{\:\,\:\mu}+T_{\mu\,\:\alpha}^{\:\:\sigma}\delta\omega_{\sigma\,\:\beta}^{\:\:\,\mu},\label{eq:ric1}
\end{equation}
where $\nabla$ acts on all the indices of $\omega$. Since $\omega$
is independent from $g$, we have:
\begin{equation}
\delta_{\omega}\mathcal{R}=\delta_{\omega}g^{\alpha\beta}R_{\alpha\beta}=g^{\alpha\beta}\delta_{\omega}R_{\alpha\beta}=g^{\alpha\beta}\left(\nabla_{\mu}\delta\omega_{\alpha\,\:\beta}^{\:\:\,\mu}-\nabla_{\alpha}\delta\omega_{\mu\,\:\beta}^{\:\:\mu}+T_{\mu\,\:\alpha}^{\:\:\sigma}\delta\omega_{\sigma\,\:\beta}^{\:\:\mu}\right).\label{eq:star}
\end{equation}
Inserting (\ref{eq:star}) to (\ref{eq:x}) gives:
\begin{equation}
\delta_{\omega}f\left(\mathcal{R}\right)=f'\left(\mathcal{R}\right)g^{\alpha\beta}\nabla_{\mu}\delta\omega_{\alpha\,\:\beta}^{\:\:\mu}-f'\left(\mathcal{R}\right)g^{\alpha\beta}\nabla_{\alpha}\delta\omega_{\mu\,\:\beta}^{\:\:\mu}+f'\left(\mathcal{R}\right)g^{\alpha\beta}T_{\mu\,\:\alpha}^{\:\:\sigma}\delta\omega_{\sigma\,\:\beta}^{\:\:\mu}.\label{huf}
\end{equation}

Let us focus on the first term of the RHS of (\ref{huf}). We could
write the following relation concerning the first term:
\begin{equation}
\nabla_{\mu}\underset{X^{\mu}}{\underbrace{f'\left(\mathcal{R}\right)g^{\alpha\beta}\delta\omega_{\alpha\,\:\beta}^{\:\:\mu}}}=\left(\nabla_{\mu}f'\left(\mathcal{R}\right)g^{\alpha\beta}\right)\delta\omega_{\alpha\,\:\beta}^{\:\:\,\mu}+f'\left(\mathcal{R}\right)g^{\alpha\beta}\nabla_{\mu}\delta\omega_{\alpha\,\:\beta}^{\:\:\,\mu},\label{eq:aaa}
\end{equation}
with:
\[
X^{\mu}=f'\left(\mathcal{R}\right)g^{\alpha\beta}\delta\omega_{\alpha\,\:\beta}^{\:\,\:\mu}.
\]
Moreover, we could have:
\begin{equation}
\nabla_{\mu}X^{\alpha}=\partial_{\mu}X^{\alpha}+\omega_{\mu\,\:\sigma}^{\:\:\alpha}X^{\sigma}=\underset{\widetilde{\nabla}_{\mu}X^{\alpha}}{\underbrace{\partial_{\mu}X^{\alpha}+\Gamma_{\mu\,\:\sigma}^{\:\:\alpha}X^{\sigma}}}+Q_{\mu\,\:\sigma}^{\:\:\,\alpha}X^{\sigma}+C_{\mu\,\:\sigma}^{\:\:\alpha}X^{\sigma},\label{eq:y}
\end{equation}
where $\widetilde{\nabla}$ is the Levi-Civita connection. $Q$ and
$C$ are the non-metricity and the torsion parts of $\omega$, known
respectively as the disformation/deflection and the contorsion tensor
\cite{nuaing}: 
\begin{align}
Q_{\mu\alpha\beta} & =-\frac{1}{2}\left(\nabla_{\mu}g_{\alpha\beta}-\nabla_{\alpha}g_{\beta\mu}+\nabla_{\beta}g_{\mu\alpha}\right),\label{eq:metricpart}\\
C_{\mu\alpha\beta} & =\frac{1}{2}\left(T_{\mu\alpha\beta}+T_{\alpha\beta\mu}-T_{\beta\mu\alpha}\right).\label{eq:connectionpart}
\end{align}
Using (\ref{eq:y}), we have:
\begin{equation}
\nabla_{\mu}X^{\mu}=\underset{\widetilde{\nabla}^{\mu}X_{\mu}}{\underbrace{\widetilde{\nabla}_{\mu}X^{\mu}}}+\left(Q_{\mu\,\:\sigma}^{\:\:\,\mu}+C_{\mu\,\:\sigma}^{\:\,\:\mu}\right)X^{\sigma},\label{eq:bbb}
\end{equation}
where we use $\widetilde{\nabla}_{\mu}X^{\mu}=\widetilde{\nabla}^{\mu}X_{\mu}$
using the fact that $\widetilde{\nabla}_{\mu}g^{\alpha\beta}=0$ for
the Levi-Civita connection. With (\ref{eq:aaa}) and (\ref{eq:bbb}),
we obtain:
\begin{equation}
f'\left(\mathcal{R}\right)g^{\alpha\beta}\nabla_{\mu}\delta\omega_{\alpha\,\:\beta}^{\:\:\mu}=\widetilde{\nabla}^{\mu}X_{\mu}+\left(Q_{\mu\,\:\sigma}^{\:\:\,\mu}+C_{\mu\,\:\sigma}^{\:\,\:\mu}\right)X^{\sigma}-\left(\nabla_{\mu}f'\left(\mathcal{R}\right)g^{\alpha\beta}\right)\delta\omega_{\alpha\,\:\beta}^{\:\:\,\mu}.\label{eq:ccc}
\end{equation}
Doing the same way with the second term in the RHS of (\ref{huf}),
we have:
\begin{equation}
f'\left(\mathcal{R}\right)g^{\alpha\beta}\nabla_{\alpha}\delta\omega_{\mu\,\:\beta}^{\:\:\mu}=\widetilde{\nabla}^{\alpha}Y_{\alpha}+\left(Q_{\alpha\,\:\sigma}^{\:\:\,\alpha}+C_{\alpha\,\:\sigma}^{\:\:\,\alpha}\right)Y^{\sigma}-\left(\nabla_{\alpha}f'\left(\mathcal{R}\right)g^{\alpha\beta}\right)\delta\omega_{\mu\,\:\beta}^{\,\:\:\mu},\label{eq:ddd}
\end{equation}
where $X$ and $Y$ satisfies:
\begin{align*}
Y^{\sigma} & =f'\left(\mathcal{R}\right)g^{\sigma\beta}\delta\omega_{\mu\,\:\beta}^{\:\,\:\mu},\\
X^{\sigma} & =f'\left(\mathcal{R}\right)g^{\alpha\beta}\delta\omega_{\alpha\,\:\beta}^{\,\:\:\sigma}.
\end{align*}
Inserting (\ref{eq:ccc}) and (\ref{eq:ddd}) to (\ref{huf}) gives:
\begin{align}
\delta_{\omega}f\left(\mathcal{R}\right)= & \widetilde{\nabla}^{\mu}X_{\mu}+\left(Q_{\mu\,\:\sigma}^{\:\:\,\mu}+C_{\mu\,\:\sigma}^{\:\,\:\mu}\right)X^{\sigma}-\left(\nabla_{\mu}f'\left(\mathcal{R}\right)g^{\alpha\beta}\right)\delta\omega_{\alpha\,\:\beta}^{\:\:\mu}\label{eq:eee}\\
 & -\widetilde{\nabla}^{\alpha}Y_{\alpha}-\left(Q_{\alpha\,\:\sigma}^{\:\:\,\alpha}+C_{\alpha\,\:\sigma}^{\:\:\,\alpha}\right)Y^{\sigma}+\left(\nabla_{\alpha}f'\left(\mathcal{R}\right)g^{\alpha\beta}\right)\delta\omega_{\mu\,\:\beta}^{\:\:\mu}+f'\left(\mathcal{R}\right)g^{\alpha\beta}T_{\mu\,\:\alpha}^{\:\:\sigma}\delta\omega_{\sigma\,\:\beta}^{\:\:\mu}.\nonumber 
\end{align}
Changing the dummy indices and performing some tensor algebra straightforwardly,
(\ref{eq:eee}) could be written as:
\begin{align}
\delta_{\omega}f\left(\mathcal{R}\right)= & \widetilde{\nabla}^{\mu}\left(X_{\mu}-Y_{\mu}\right)+\left(\left(Q_{\lambda\,\:\nu}^{\:\:\lambda}+C_{\lambda\,\:\nu}^{\:\:\lambda}-\nabla_{\nu}\right)2\delta_{\mu}^{\:\left[\nu\right.}g^{\left.\alpha\right]\beta}+T_{\mu}^{\:\:\alpha\beta}\right)f'\left(\mathcal{R}\right)\delta\omega_{\alpha\,\:\beta}^{\:\:\mu}.\label{eq:fff}
\end{align}

Let us keep the result for the moment and focus to the remaining terms
in relation (\ref{eq:dS-1}). The third term in the RHS of (\ref{eq:dS-1})
is the variation of the matter Lagrangian with respect to the connection:
\begin{equation}
\delta_{\omega}S_{\mathrm{matter}}\left[g,\omega\right]:=-\intop_{\mathcal{M}}\kappa\mathcal{H}_{\:\:\mu}^{\alpha\,\:\beta}\delta\omega_{\alpha\,\:\beta}^{\:\:\,\mu}\mathrm{vol}.\label{eq:ggg}
\end{equation}
 $\mathcal{H}$ is known as the hypermomentum, which is zero for standard
matter without internal degrees of freedom such as spins \cite{Hehl2,Hehl3,Hehl4,Hehl5}.

The next term in the RHS of (\ref{eq:dS-1}) is the $S_{LM}$ part:
\begin{equation}
\delta_{\omega}S_{LM}=\intop_{\mathcal{M}}\delta_{\omega}\left(\chi^{\mu}T_{\alpha\;\:\mu}^{\;\alpha}\mathrm{vol}\right)=\intop_{\mathcal{M}}\underset{0}{\underbrace{\left(\delta_{\omega}\chi^{\mu}\right)}}T_{\alpha\;\:\mu}^{\;\:\alpha}\mathrm{vol}+\intop_{\mathcal{M}}\chi^{\mu}\left(\delta_{\omega}T_{\alpha\;\:\mu}^{\;\:\alpha}\right)\mathrm{vol}+\intop_{\mathcal{M}}\chi^{\mu}T_{\alpha\;\:\mu}^{\;\:\alpha}\underset{0}{\underbrace{\left(\delta_{\omega}\mathrm{vol}\right)}},\label{eq:z}
\end{equation}
since $\chi$ and ($\mathrm{vol}$) are independent of $\omega$.
One could easily obtain $\delta_{\omega}T_{\alpha\;\:\mu}^{\:\;\alpha}$
from the definition of torsion tensor in (\ref{eq:definitiontorsion}):
\begin{equation}
\delta_{\omega}T_{\nu\;\:\sigma}^{\;\nu}=\left(\delta_{\mu}^{\alpha}\delta_{\sigma}^{\beta}-\delta_{\sigma}^{\alpha}\delta_{\mu}^{\beta}\right)\delta\omega_{\alpha\,\:\beta}^{\:\:\mu},\label{eq:z1}
\end{equation}
which could be inserted to (\ref{eq:z}) to give:
\begin{equation}
\delta_{\omega}S_{LM}=\intop_{\mathcal{M}}\left(\chi^{\beta}\delta_{\mu}^{\alpha}-\chi^{\alpha}\delta_{\mu}^{\beta}\right)\delta\omega_{\alpha\,\:\beta}^{\:\:\mu}\mathrm{vol}.\label{eq:hhh}
\end{equation}

The last term in the RHS of (\ref{eq:dS-1}) is $S_{\mathrm{sym}}$:
\begin{equation}
\delta_{\omega}S_{\mathrm{sym}}=\intop_{\mathcal{M}}\delta_{\omega}\left(\zeta_{\alpha\beta}g^{\left[\alpha\beta\right]}\mathrm{vol}\right)=0,\label{eq:iii}
\end{equation}
since $\zeta$, $g$ and ($\mathrm{vol}$) are independent of $\omega$. 

Inserting (\ref{eq:fff}), (\ref{eq:ggg}), (\ref{eq:hhh}), and (\ref{eq:iii})
to (\ref{eq:dS-1}) gives:
\begin{equation}
\delta_{\omega}S=\intop_{\mathcal{M}}\widetilde{\nabla}^{\mu}\left(X_{\mu}-Y_{\mu}\right)\mathrm{vol}+\intop_{\mathcal{M}}\left(\left(\left(Q_{\lambda\,\:\nu}^{\:\:\lambda}+C_{\lambda\,\:\nu}^{\:\:\lambda}-\nabla_{\nu}\right)2\delta_{\mu}^{\:\left[\nu\right.}g^{\left.\alpha\right]\beta}+T_{\mu}^{\:\:\alpha\beta}\right)f'\left(\mathcal{R}\right)-\kappa\mathcal{H}_{\:\:\mu}^{\alpha\,\:\beta}-\left(\chi^{\alpha}\delta_{\mu}^{\beta}-\chi^{\beta}\delta_{\mu}^{\alpha}\right)\right)\delta\omega_{\alpha\,\:\beta}^{\:\:\mu}\mathrm{vol}.\label{eq:z2}
\end{equation}
The first term in the RHS of (\ref{eq:z2}) is a total divergence,
which is equal to zero if $\mathcal{M}$ is compact \cite{Baez};
let us apply this case for simplicity. Then one is left with the second,
third and last terms. The Euler-Lagrange equation which minimize the
action (\ref{eq:MAfR}) for any variation of the connection is then:
\begin{equation}
\left(\left(Q_{\lambda\,\:\nu}^{\:\:\lambda}+C_{\lambda\,\:\nu}^{\:\:\lambda}-\nabla_{\nu}\right)\left(2\delta_{\mu}^{\:\left[\nu\right.}g^{\left.\alpha\right]\beta}\right)+T_{\mu}^{\:\:\alpha\beta}\right)f'\left(\mathcal{R}\right)=\kappa\mathcal{H}_{\:\:\mu}^{\alpha\,\:\beta}+\chi^{\alpha}\delta_{\mu}^{\beta}-\chi^{\beta}\delta_{\mu}^{\alpha}.\label{eq:jjj}
\end{equation}
Notice that the covariant derivative $\nabla_{\nu}$ acts on both
$\delta_{\mu}^{\:\left[\nu\right.}g^{\left.\alpha\right]\beta}$ and
$f'\left(\mathcal{R}\right)$. Moreover, using (\ref{eq:metricpart})
and (\ref{eq:connectionpart}), and using the symmetries of $Q$ and
$C$ \cite{nuaing}, (\ref{eq:jjj}) becomes:
\begin{equation}
\underset{P_{\mu}^{\:\:\alpha\beta}}{\underbrace{\left(\left(T_{\lambda\,\:\nu}^{\:\:\lambda}-\frac{1}{2}g^{\sigma\lambda}\nabla_{\nu}g_{\sigma\lambda}-\nabla_{\nu}\right)\left(2\delta_{\mu}^{\:\left[\nu\right.}g^{\left.\alpha\right]\beta}\right)+T_{\mu}^{\:\:\alpha\beta}\right)}}f'\left(\mathcal{R}\right)=\kappa\mathcal{H}_{\:\:\mu}^{\alpha\,\:\beta}+2\chi^{\left[\alpha\right.}\delta_{\mu}^{\left.\beta\right]}.\label{eq:EL2}
\end{equation}
$\boldsymbol{P}$ is known as the Palatini tensor, and (\ref{eq:EL2})
is the hypermomentum equation of the Metric-Affine $f\left(\mathcal{R}\right)$-Gravity.

One could show that (\ref{eq:EL2}) is equivalent with the original
EL equation in the standard derivation in \cite{Olmo,Sotiriou} as
follows. Breaking the antisymmetric part explicitly into its component,
(\ref{eq:EL2}) could be written as:

\begin{align}
\left(\left(-\frac{1}{2}\left(g^{\sigma\lambda}\nabla_{\mu}g_{\sigma\lambda}\right)g^{\alpha\beta}-\nabla_{\mu}g^{\alpha\beta}\right)+\delta_{\mu}^{\alpha}\left(\frac{1}{2}\left(g^{\sigma\lambda}\nabla_{\nu}g_{\sigma\lambda}\right)g^{\nu\beta}+\nabla_{\nu}g^{\nu\beta}\right)\right)f'\left(\mathcal{R}\right)\label{eq:bongkar}\\
+\left(T_{\lambda\,\:\mu}^{\:\:\lambda}g^{\alpha\beta}-T_{\lambda}^{\:\:\lambda\beta}\delta_{\mu}^{\alpha}+T_{\mu}^{\:\:\alpha\beta}\right)f'\left(\mathcal{R}\right) & =\kappa\mathcal{H}_{\:\:\mu}^{\alpha\,\:\beta}+\chi^{\alpha}\delta_{\mu}^{\beta}-\chi^{\beta}\delta_{\mu}^{\alpha}\nonumber 
\end{align}
Using the fact that $\nabla_{\mu}\mathfrak{g}=\mathfrak{g}g^{\alpha\beta}\nabla_{\mu}g_{\alpha\beta}$,
where $\mathfrak{g}$ is the determinant of $g_{\alpha\beta}$, one
could obtain:
\begin{equation}
\frac{1}{\sqrt{-\mathfrak{g}}}\nabla_{\mu}\sqrt{-\mathfrak{g}}g^{\alpha\beta}=\left(\frac{1}{2}g^{\sigma\lambda}\nabla_{\mu}g_{\sigma\lambda}+\nabla_{\mu}\right)g^{\alpha\beta}.\label{eq:xx}
\end{equation}
Inserting (\ref{eq:xx}) to (\ref{eq:bongkar}) gives:
\[
\left(-\left(\nabla_{\mu}\sqrt{-\mathfrak{g}}g^{\alpha\beta}\right)+\delta_{\mu}^{\alpha}\left(\nabla_{\nu}\sqrt{-\mathfrak{g}}g^{\nu\beta}\right)+\sqrt{-\mathfrak{g}}T_{\lambda\,\:\mu}^{\:\:\lambda}g^{\alpha\beta}-\sqrt{-\mathfrak{g}}T_{\lambda}^{\:\:\lambda\beta}\delta_{\mu}^{\alpha}+\sqrt{-\mathfrak{g}}T_{\mu}^{\:\:\alpha\beta}\right)f'\left(\mathcal{R}\right)=\sqrt{-\mathfrak{g}}\left(\kappa\mathcal{H}_{\:\:\mu}^{\alpha\,\:\beta}+2\chi^{\left[\alpha\right.}\delta_{\mu}^{\left.\beta\right]}\right),
\]
which is equivalent to the standard form of hypermomentum equation
in \cite{Olmo,Sotiriou}, up to the factor 2, due to the difference
in the definition of torsion tensor (\ref{eq:definitiontorsion})
(to be specific, the definition of torsion in \cite{Olmo,Sotiriou}
is $T_{\mu\,\:\beta}^{\:\:\alpha}=\frac{1}{2}\left(\omega_{\mu\,\:\beta}^{\:\:\alpha}-\omega_{\beta\,\:\mu}^{\:\:\alpha}\right)$).

\subsubsection*{The Traceless Torsion and Symmetricity Constraints}

There are 2 remaining equations of motion that arise from the variation
of $S$ with respect to the Lagrange multipliers. The variation of
$S$ with respect to $\chi$ gives the traceless torsion constraint
$T_{\mu}=T_{\lambda\,\:\mu}^{\:\:\lambda}=0.$ It should be kept in
mind, that this equation does not emerge from the dynamics, but it
is introduced in the kinematical level, in order to remove the inconsistencies
arise under the projective invariant transformation \cite{Iosifidis,Sotiriou,Iosifidis4}.
Notice that taking the torsion to be traceless is not the only possible
way to cure the inconsistencies, one could as well take the Weyl tensor
$Q_{\mu}$ (of the non-metricity factor), or the linear combination
of $T_{\mu}$ and $Q_{\mu}$, to be zero. See \cite{Iosifidis,Iosifidis4}.

The last equation of motion is obtained from variation of $S$ with
respect to $\zeta$, this gives the symmetricity of the metric  $g^{\left[\alpha\beta\right]}=0.$
Similar with the previous constraint, this equation does not emerge
from the dynamics, but it is introduced to guarantee the consistency
of the equations of motion that arise under an \textit{a priori} assumption
that the metric in Metric-Affine $f\left(\mathcal{R}\right)$-gravity
is always symmetric.

To conclude this section, the variation of $S$ with respect to $g,$
$\omega,$ $\chi$, and $\zeta$, gives 4 equations of motion:
\begin{align}
f'\left(\mathcal{R}\right)R_{\alpha\beta}-\frac{1}{2}\left(f\left(\mathcal{R}\right)+\chi^{\mu}T_{\lambda\,\:\mu}^{\:\:\lambda}+\zeta_{\mu\nu}g^{\left[\mu\nu\right]}\right)g_{\alpha\beta}+\zeta_{\left[\alpha\beta\right]} & =\mathcal{\kappa T}_{\alpha\beta},\label{eq:ELmafr1}\\
\left(\left(T_{\lambda\,\:\nu}^{\:\:\lambda}+Q_{\lambda\,\:\nu}^{\:\:\lambda}-\nabla_{\nu}\right)\left(2\delta_{\mu}^{\:\left[\nu\right.}g^{\left.\alpha\right]\beta}\right)+T_{\mu}^{\:\:\alpha\beta}\right)f'\left(\mathcal{R}\right) & =\kappa\mathcal{H}_{\:\:\mu}^{\alpha\,\:\beta}+2\chi^{\left[\alpha\right.}\delta_{\mu}^{\left.\beta\right]},\label{eq:ELmafr2}\\
T_{\lambda\,\:\mu}^{\:\:\lambda} & =0,\label{eq:ELmaffr3}\\
g^{\left[\alpha\beta\right]} & =0.\label{eq:ELmafr4}
\end{align}
Let us solve (\ref{eq:ELmafr2}) for $\chi$, following \cite{Sotiriou}.
Contracting the indices $(\beta,\mu)$ gives $\chi^{\alpha}=-\frac{1}{3}\kappa\mathcal{H}_{\:\:\beta}^{\alpha\,\:\beta}.$
On the other hand, from the symmetries of (\ref{eq:ELmafr1}), we
could deduce that $\zeta_{\left[\alpha\beta\right]}\equiv-R_{\left[\alpha\beta\right]}$,
since $g_{\alpha\beta}$ and $\mathcal{T}_{\alpha\beta}$ are always
symmetric. Inserting these results together with (\ref{eq:ELmaffr3})-(\ref{eq:ELmafr4})
to (\ref{eq:ELmafr1})-(\ref{eq:ELmafr2}) gives the simplified version
of the equations of motion:
\begin{align}
f'\left(\mathcal{R}\right)R_{\left(\alpha\beta\right)}-\frac{1}{2}f\left(\mathcal{R}\right)g_{\alpha\beta} & =\mathcal{\kappa T}_{\alpha\beta},\label{eq:ELmafr1-1}\\
\left(\left(Q_{\lambda\,\:\nu}^{\:\:\lambda}-\nabla_{\nu}\right)\left(2\delta_{\mu}^{\:\left[\nu\right.}g^{\left.\alpha\right]\beta}\right)+T_{\mu}^{\:\:\alpha\beta}\right)f'\left(\mathcal{R}\right) & =\kappa\mathcal{H}_{\:\:\mu}^{\alpha\,\:\beta}-\frac{2}{3}\kappa\mathcal{H}_{\:\:\sigma}^{\left[\alpha\right|\,\:\sigma}\delta_{\mu}^{\left|\beta\right]},\label{eq:ELmafr2-1}\\
T_{\lambda\,\:\mu}^{\:\:\lambda} & =0,\label{eq:ELmaffr3-1}
\end{align}
where we omit the last equation $g^{\left[\alpha\beta\right]}=0,$
since we are always working with a symmetric metric $g$.

\subsubsection*{Special Case: Metric-Affine GR (MAGR)}

The theory of Metric-Affine General Relativity (MAGR) (or Generalized
Palatini gravity) is a special case of Metric-Affine $f\left(\mathcal{R}\right)$-Gravity
for $f\left(\mathcal{R}\right)=\mathcal{R}$. With this requirement,
(\ref{eq:ELmafr1-1})-(\ref{eq:ELmaffr3-1}) becomes:
\begin{align}
R_{\left(\alpha\beta\right)}-\frac{1}{2}\mathcal{R}g_{\alpha\beta} & =\mathcal{\kappa T}_{\alpha\beta},\label{eq:EoM1}\\
\left(Q_{\lambda\,\:\nu}^{\:\:\lambda}-\nabla_{\nu}\right)\left(2\delta_{\mu}^{\:\left[\nu\right.}g^{\left.\alpha\right]\beta}\right)+T_{\mu}^{\:\:\alpha\beta} & =\kappa\left(\mathcal{H}_{\:\:\mu}^{\alpha\,\:\beta}-\frac{2}{3}\mathcal{H}_{\:\:\;\;\,\sigma}^{\left[\alpha\right|\,\:\sigma}\delta_{\mu}^{\left|\beta\right]}\right),\label{eq:EoM2}\\
T_{\lambda\,\:\mu}^{\:\:\lambda} & =0.\label{eq:EoM3}
\end{align}
As the (3+1) decomposition of the stress-energy-momentum equation
(\ref{eq:EoM1}) had been done in \cite{nuaing}, in the next part
of this article, we will focus on the (3+1) decomposition of the hypermomentum
equation (\ref{eq:EoM2}).

\section{(3+1) Hypermomentum Equation}

\subsubsection*{(3+1) Decomposition of Hypermomentum}

One could write the hypermomentum (\ref{eq:ggg}) with the spacetime
index $\alpha$ is hidden as follows \cite{nuaing}:
\begin{equation}
\mathcal{H}\left(X,Y^{*}\right):=X^{\mu}Y_{\beta}\mathcal{H}_{\:\:\mu}^{\alpha\,\:\beta}\partial_{\alpha},\qquad X\in T_{p}\mathcal{M},\;Y^{*}\in T_{p}^{*}\mathcal{M},\label{eq:efh}
\end{equation}
where $\mathcal{H}\left(\partial_{\mu},dx^{\beta}\right):=\mathcal{H}_{\:\:\mu}^{\alpha\,\:\beta}\partial_{\alpha}.$
Adapting the terminology of gauge theory, let us called $\alpha$
as the spacetime index, and $\left(\mu,\beta\right)$ as the 'internal'
indices. With this notation, the hypermomentum could be decomposed
by its 'internal' indices into the normal and parallel parts, with
respect to the hypersurface $\Sigma$ \cite{nuaing}:
\begin{align}
\mathcal{H}\left(\hat{n},\hat{n}^{*}\right) & =-\left\langle \hat{n}^{*},\mathcal{H}\left(\hat{n},\hat{n}^{*}\right)\right\rangle \hat{n}+\left\langle \,^{3}dx^{i},\mathcal{H}\left(\hat{n},\hat{n}^{*}\right)\right\rangle \partial_{i},\nonumber \\
\mathcal{H}\left(\hat{n},\,^{3}dx^{i}\right) & =-\left\langle \hat{n}^{*},\mathcal{H}\left(\hat{n},\,^{3}dx^{i}\right)\right\rangle \hat{n}+\left\langle \,^{3}dx^{j},\mathcal{H}\left(\hat{n},\,^{3}dx^{i}\right)\right\rangle \partial_{j},\nonumber \\
\mathcal{H}\left(\partial_{i},\hat{n}^{*}\right) & =-\left\langle \hat{n}^{*},\mathcal{H}\left(\partial_{i},\hat{n}^{*}\right)\right\rangle \hat{n}+\left\langle \,^{3}dx^{j},\mathcal{H}\left(\partial_{i},\hat{n}^{*}\right)\right\rangle \partial_{j},\label{eq:h1}\\
\mathcal{H}\left(\partial_{i},\,^{3}dx^{j}\right) & =-\left\langle \hat{n}^{*},\mathcal{H}\left(\partial_{i},\,^{3}dx^{j}\right)\right\rangle \hat{n}+\left\langle \,^{3}dx^{k},\mathcal{H}\left(\partial_{i},\,^{3}dx^{j}\right)\right\rangle \partial_{k}.\nonumber 
\end{align}
This is similar to the decomposition of the stress-energy-momentum
tensor $\mathcal{T}$ into its 3 components, i.e., the energy $E$,
the momentum $p_{i}$, and the stress $\mathcal{S}_{ij}$, where the
decomposition is based on the split of the 'spacetime' into space
and time. Another distinct decomposition of $\mathcal{H}$, which
is based on the symmetricity of the indices gives the spin, shear,
and dilation parts of $\mathcal{H}$ as introduced in \cite{Hehl,Hehl2,Hehl3,Hehl4,Hehl5}.
In this article, we will apply both of these decompositions.

\subsubsection*{The Dilation, Shear, and Spin (Rotation) Parts of the Hypermomentum}

As proposed in \cite{Hehl}, the hypermomentum could be decomposed
according to the symmetries of its 'internal' indices as $\mathcal{H}_{\:\:\alpha\beta}^{\mu}=\mathcal{H}_{\:\:\left(\alpha\beta\right)}^{\mu}+\mathcal{H}_{\:\:\left[\alpha\beta\right]}^{\mu}.$
The antisymetric part is defined as the spin (rotation) of the hypermomentum:
\[
\mathcal{H}_{\:\:\left[\alpha\beta\right]}^{\mu}:=\Omega_{\:\:\alpha\beta}^{\mu},
\]
while the symmetric part could be furthermore decomposed as $\mathcal{H}_{\:\:\left(\alpha\beta\right)}^{\mu}=\frac{1}{d}g_{\alpha\beta}\mathcal{D}^{\mu}+\Lambda_{\:\:\alpha\beta}^{\mu},$
with $d$ is the dimension of the manifold. $\mathcal{D}^{\mu}$ is
the dilation vector:
\[
\mathcal{D}^{\mu}:=\mathrm{tr}\mathcal{H}^{\mu}=\mathcal{H}_{\:\:\nu}^{\mu\,\:\nu},
\]
and $\Lambda_{\:\:\alpha\beta}^{\mu}$ as the shear of the hypermomentum:
\[
\Lambda_{\:\:\alpha\beta}^{\mu}:=\mathcal{H}_{\:\:\left(\alpha\beta\right)}^{\mu}-\frac{1}{d}g_{\alpha\beta}\mathcal{H}_{\:\:\nu}^{\mu\,\:\nu}.
\]

Notice that these decompositions is not defined for $\mathcal{H}_{\:\:\alpha}^{\mu\,\:\beta},$
but instead for $\mathcal{H}_{\:\:\alpha\beta}^{\mu}$ with the last
index lowered. Hence, let us define $\mathcal{H}^{*}$ as follows:
\begin{equation}
\mathcal{H}\left(Y,Z^{*}\right)=\mathcal{H}\left(Y,g\left(Z\right)\right)=\mathcal{H}^{*}\left(Y,Z\right),\qquad Z\in T_{p}\mathcal{M}.\label{eq:p0}
\end{equation}
With this, then one could obtain:
\begin{equation}
\left\langle X^{*},\mathcal{H}\left(Y,Z^{*}\right)\right\rangle =\left\langle X^{*},\mathcal{H}^{*}\left(Y,Z\right)\right\rangle =\frac{1}{4}\left\langle X^{*},\boldsymbol{\mathcal{D}}\right\rangle g\left(Y,Z\right)+\left\langle X^{*},\Lambda\left(Y,Z\right)\right\rangle +\left\langle X^{*},\Omega\left(Y,Z\right)\right\rangle ,\label{eq:pentings}
\end{equation}
or in indices:
\[
\left\langle X^{*},\mathcal{H}\left(Y,Z^{*}\right)\right\rangle =Y^{\alpha}Z_{\beta}\mathcal{H}_{\:\:\alpha}^{\mu\,\:\beta}X_{\mu}=X_{\mu}Y^{\alpha}Z^{\beta}\mathcal{H}_{\:\:\alpha\beta}^{\mu}=X_{\mu}Y^{\alpha}Z^{\beta}\left(\frac{1}{4}g_{\alpha\beta}\mathcal{D}^{\mu}+\Lambda_{\:\:\alpha\beta}^{\mu}+\Omega_{\:\:\alpha\beta}^{\mu}\right).
\]
The shear $\Lambda$ and the rotation $\Omega$ satisfy the following
symmetries:
\begin{align}
\Lambda\left(Y,Z\right) & =\Lambda\left(Z,Y\right),\label{eq:p2}\\
\Omega\left(Y,Z\right) & =-\Omega\left(Z,Y\right).\label{eq:p1}
\end{align}
One could show that:
\begin{equation}
\left\langle X^{*},\mathcal{H}\left(Y,\,^{3}dx^{i}\right)\right\rangle =\left\langle X^{*},\,^{3}q^{ij}\mathcal{H}^{*}\left(Y,\partial_{j}\right)\right\rangle ,\label{eq:p00}
\end{equation}
which will be used in the following derivation.

Now using (\ref{eq:pentings}) and (\ref{eq:p00}), together with
the help of (\ref{eq:p0}) and (\ref{eq:p2})-(\ref{eq:p1}), the
temporal and spatial components in (\ref{eq:h1}) becomes:
\begin{align}
\left\langle \hat{n}^{*},\mathcal{H}\left(\hat{n},\hat{n}^{*}\right)\right\rangle  & =-\frac{1}{4}\left\langle \hat{n}^{*},\boldsymbol{\mathcal{D}}\right\rangle +\left\langle \hat{n}^{*},\Lambda\left(\hat{n},\hat{n}\right)\right\rangle ,\label{eq:s1}\\
\left\langle \,^{3}dx^{i},\mathcal{H}\left(\hat{n},\hat{n}^{*}\right)\right\rangle  & =-\frac{1}{4}\left\langle \,^{3}dx^{i},\boldsymbol{\mathcal{D}}\right\rangle +\left\langle \,^{3}dx^{i},\Lambda\left(\hat{n},\hat{n}\right)\right\rangle ,\nonumber 
\end{align}
\begin{align}
\left\langle \hat{n}^{*},\mathcal{H}\left(\hat{n},\,^{3}dx^{i}\right)\right\rangle  & =\,^{3}q^{ij}\left(\left\langle \hat{n}^{*},\Lambda\left(\hat{n},\partial_{j}\right)\right\rangle +\left\langle \hat{n}^{*},\Omega\left(\hat{n},\partial_{j}\right)\right\rangle \right),\label{eq:s2}\\
\left\langle \,^{3}dx^{i},\mathcal{H}\left(\hat{n},\,^{3}dx^{j}\right)\right\rangle  & =\,^{3}q^{jk}\left(\left\langle \,^{3}dx^{i},\Lambda\left(\hat{n},\partial_{k}\right)\right\rangle +\left\langle \,^{3}dx^{i},\Omega\left(\hat{n},\partial_{k}\right)\right\rangle \right),\nonumber 
\end{align}
\begin{align}
\left\langle \hat{n}^{*},\mathcal{H}\left(\partial_{i},\hat{n}^{*}\right)\right\rangle  & =\left\langle \hat{n}^{*},\Lambda\left(\hat{n},\partial_{i}\right)\right\rangle -\left\langle \hat{n}^{*},\Omega\left(\hat{n},\partial_{i}\right)\right\rangle ,\label{eq:s3}\\
\left\langle \,^{3}dx^{i},\mathcal{H}\left(\partial_{j},\hat{n}^{*}\right)\right\rangle  & =\left\langle \,^{3}dx^{i},\Lambda\left(\hat{n},\partial_{j}\right)\right\rangle -\left\langle \,^{3}dx^{i},\Omega\left(\hat{n},\partial_{j}\right)\right\rangle ,\nonumber 
\end{align}
\begin{align}
\left\langle \hat{n}^{*},\mathcal{H}\left(\partial_{i},\,^{3}dx^{j}\right)\right\rangle  & =\frac{1}{4}\delta_{i}^{j}\left\langle \hat{n}^{*},\boldsymbol{\mathcal{D}}\right\rangle +\,^{3}q^{jk}\left(\left\langle \hat{n}^{*},\Lambda\left(\partial_{i},\partial_{k}\right)\right\rangle +\left\langle \hat{n}^{*},\Omega\left(\partial_{i},\partial_{k}\right)\right\rangle \right),\label{eq:s4}\\
\left\langle \,^{3}dx^{i},\mathcal{H}\left(\partial_{j},\,^{3}dx^{k}\right)\right\rangle  & =\frac{1}{4}\delta_{j}^{k}\left\langle \,^{3}dx^{i},\boldsymbol{\mathcal{D}}\right\rangle +\,^{3}q^{kl}\left(\left\langle \,^{3}dx^{i},\Lambda\left(\partial_{j},\partial_{l}\right)\right\rangle +\left\langle \,^{3}dx^{i},\Omega\left(\partial_{j},\partial_{l}\right)\right\rangle \right).\nonumber 
\end{align}
Notice that by the symmetries (\ref{eq:p2})-(\ref{eq:p1}), we could
simplify (\ref{eq:s1})-(\ref{eq:s4}) in writing them with 10 hypersurface
variables  that construct a complete hypermomentum tensor as follows.
The dilation $\boldsymbol{\mathcal{D}}$ is a 4-vector, hence it could
be decomposed into the (scalar) temporal dilation $\left\langle \hat{n}^{*},\boldsymbol{\mathcal{D}}\right\rangle $
and (3-vector) spatial dilation $\left\langle \,^{3}dx^{i},\boldsymbol{\mathcal{D}}\right\rangle $:
\[
\boldsymbol{\mathcal{D}}=\underset{\mathcal{D}\left(\hat{n}\right)}{\underbrace{-\left\langle \hat{n}^{*},\boldsymbol{\mathcal{D}}\right\rangle }}\hat{n}+\underset{\mathcal{D}^{i}}{\underbrace{\left\langle \,^{3}dx^{i},\boldsymbol{\mathcal{D}}\right\rangle }}\partial_{i}.
\]
We usually write $\mathcal{D}^{i}\partial_{i}$ as $\,^{3}\mathcal{\boldsymbol{D}}$.
Hence, the 4-vector dilation $\boldsymbol{\mathcal{D}}$ could be
constructed from 2 hypersurface variables, namely, the scalar $\mathcal{D}\left(\hat{n}\right)$
and 3-vector $\,^{3}\mathcal{\boldsymbol{D}}$.

The shear $\boldsymbol{\Lambda}$, which is a traceless, symmetric
matrix-valued 4-vector, could be decomposed into:
\begin{align}
\Lambda\left(\hat{n},\hat{n}\right) & =\textrm{tr}\boldsymbol{\sigma}=\,^{3}q^{ij}\boldsymbol{\sigma}_{ij},\label{eq:shear1}\\
\Lambda\left(\hat{n},\partial_{i}\right) & =\Lambda\left(\partial_{i},\hat{n}\right):=\boldsymbol{\tau}_{i},\label{eq:shear2}\\
\Lambda\left(\partial_{i},\partial_{j}\right) & =\Lambda\left(\partial_{j},\partial_{i}\right):=\boldsymbol{\sigma}_{ij}.\label{eq:shear3}
\end{align}
The condition (\ref{eq:shear1}) could be obtained from the tracelessness
of $\Lambda$:
\[
\textrm{tr}\Lambda=g^{*}\left(\hat{n}^{*},\hat{n}^{*}\right)\Lambda\left(\hat{n},\hat{n}\right)+2g^{*}\left(\hat{n}^{*},\,^{3}dx^{i}\right)\Lambda\left(\hat{n},\partial_{i}\right)+g^{*}\left(\,^{3}dx^{i},\,^{3}dx^{j}\right)\Lambda\left(\partial_{i},\partial_{j}\right).
\]
With definition (\ref{eq:shear3}) and using the fact that $g^{*}\left(\hat{n}^{*},\hat{n}^{*}\right)=-1$,
$g^{*}\left(\hat{n}^{*},\,^{3}dx^{i}\right)=0,$ and $g^{*}\left(\,^{3}dx^{i},\,^{3}dx^{j}\right)=\,^{3}q^{ij}$,
one could obtained (\ref{eq:shear1}). Note that $\boldsymbol{\tau}$
and $\boldsymbol{\sigma}$ are respectively, a vector-valued and matrix-valued
4-vector, therefore, they could be furthermore decomposed as follows:
\begin{align*}
\Lambda\left(\hat{n},\hat{n}\right) & =\textrm{tr}\boldsymbol{\sigma}=-\left\langle \hat{n}^{*},\textrm{tr}\boldsymbol{\sigma}\right\rangle \hat{n}+\underset{\left(\textrm{tr}\sigma\right)^{i}}{\underbrace{\left\langle \,^{3}dx^{i},\textrm{tr}\boldsymbol{\sigma}\right\rangle }}\partial_{i},\\
\Lambda\left(\hat{n},\partial_{i}\right) & =\Lambda\left(\partial_{i},\hat{n}\right)=\boldsymbol{\tau}_{i}=-\left\langle \hat{n}^{*},\boldsymbol{\tau}_{i}\right\rangle \hat{n}+\underset{\tau_{\;i}^{j}}{\underbrace{\left\langle \,^{3}dx^{j},\boldsymbol{\tau}_{i}\right\rangle }}\partial_{j},\\
\Lambda\left(\partial_{i},\partial_{j}\right) & =\Lambda\left(\partial_{j},\partial_{i}\right)=\boldsymbol{\sigma}_{ij}=-\left\langle \hat{n}^{*},\boldsymbol{\sigma}_{ij}\right\rangle \hat{n}+\underset{\mathcal{\sigma}_{\:\:ij}^{k}}{\underbrace{\left\langle \,^{3}dx^{k},\boldsymbol{\sigma}_{ij}\right\rangle }}\partial_{k}.
\end{align*}
 $\mathcal{\sigma}_{\:\:ij}^{k}$ is known as the 3-dimensional spatial
shear tensor. The shear tensor $\boldsymbol{\Lambda}$ could be constructed
from 4 hypersurface variables, namely, the 3-vector $\left\langle \hat{n}^{*},\boldsymbol{\tau}_{i}\right\rangle ,$
the matrices $\tau_{\;i}^{j}$ and $\left\langle \hat{n}^{*},\boldsymbol{\sigma}_{ij}\right\rangle ,$
and the $\tbinom{1}{2}$-tensor $\mathcal{\sigma}_{\:\:ij}^{k}$.
Note that $\left\langle \hat{n}^{*},\textrm{tr}\boldsymbol{\sigma}\right\rangle $
and $\left(\textrm{tr}\sigma\right)^{i}$ could be obtained by contracting
$\left\langle \hat{n}^{*},\boldsymbol{\sigma}_{ij}\right\rangle $
and $\left(\textrm{tr}\sigma\right)^{i}$ with $\,^{3}q^{ij}$.

Finally, the spin $\boldsymbol{\Omega}$ , which is an antisymmetric-valued
4-vector, is decomposed into:
\begin{align}
\Omega\left(\hat{n},\hat{n}\right) & =0,\label{eq:rot1}\\
\Omega\left(\hat{n},\partial_{i}\right) & =-\Omega\left(\partial_{i},\hat{n}\right)=\boldsymbol{k}_{i},\label{eq:rot2}\\
\Omega\left(\partial_{i},\partial_{j}\right) & =-\Omega\left(\partial_{j},\partial_{i}\right)=\varepsilon_{ijk}\boldsymbol{l}^{k}.\label{eq:rot3}
\end{align}
Condition (\ref{eq:rot1}) is a consequence of the antisymmetricity
of $\Omega$. $\boldsymbol{k},$ which is a vector-valued 4-vector,
is usually known as the 'boost' part of a 4-dimensional rotation,
while $\boldsymbol{l}^{k}$, also a vector valued 4-vector, is usually
known as the 3D rotation part. They could be furthermore decomposed
into:
\begin{align*}
\Omega\left(\hat{n},\partial_{i}\right) & =-\Omega\left(\partial_{i},\hat{n}\right)=\boldsymbol{k}_{i}=-\left\langle \hat{n}^{*},\boldsymbol{k}_{i}\right\rangle \hat{n}+\underset{k_{\;i}^{j}}{\underbrace{\left\langle \,^{3}dx^{j},\boldsymbol{k}_{i}\right\rangle }}\partial_{j},\\
\Omega\left(\partial_{i},\partial_{j}\right) & =-\Omega\left(\partial_{j},\partial_{i}\right)=\varepsilon_{ijk}\boldsymbol{l}^{k}=-\varepsilon_{ijk}\left\langle \hat{n}^{*},\boldsymbol{l}^{k}\right\rangle \hat{n}+\varepsilon_{ijk}\underset{l^{mk}}{\underbrace{\left\langle \,^{3}dx^{m},\boldsymbol{l}^{k}\right\rangle }}\partial_{m}.
\end{align*}
The rotation bivector $\boldsymbol{\Omega}$ then could be constructed
from 4 hypersurface variables, namely, the 3-vectors $\left\langle \hat{n}^{*},\boldsymbol{k}_{i}\right\rangle $
and $\left\langle \hat{n}^{*},\boldsymbol{l}^{k}\right\rangle $,
and the matrices $k_{\;i}^{j}$ and $l^{mk}$. Notice that the hypermomentum
$\mathcal{H}$ could be constructed completely from these 10 hypersurface
variables: $\boldsymbol{\mathcal{D}}=\left(\mathcal{D}\left(\hat{n}\right),\,^{3}\mathcal{D}\right)$,
$\boldsymbol{\tau}_{i}=\left(-\left\langle \hat{n}^{*},\boldsymbol{\tau}_{i}\right\rangle ,\tau_{\;i}^{j}\right)$,
$\boldsymbol{\sigma}_{ij}=\left(-\left\langle \hat{n}^{*},\boldsymbol{\sigma}_{ij}\right\rangle ,\mathcal{\sigma}_{\:\:ij}^{k}\right),$
$\boldsymbol{k}_{i}=\left(-\left\langle \hat{n}^{*},\boldsymbol{k}_{i}\right\rangle ,k_{\;i}^{j}\right)$,
and $\boldsymbol{l}^{k}=\left(-\left\langle \hat{n}^{*},\boldsymbol{l}^{k}\right\rangle ,l^{mk}\right)$,
and relation (\ref{eq:s1})-(\ref{eq:s4}) could be written as:
\begin{align}
\left\langle \hat{n}^{*},\mathcal{H}\left(\hat{n},\hat{n}^{*}\right)\right\rangle  & =\frac{1}{4}\mathcal{D}\left(\hat{n}\right)+\left\langle \hat{n}^{*},\textrm{tr}\boldsymbol{\sigma}\right\rangle ,\label{eq:s1-1}\\
\left\langle \,^{3}dx^{i},\mathcal{H}\left(\hat{n},\hat{n}^{*}\right)\right\rangle  & =-\frac{1}{4}\mathcal{D}^{i}+\left(\textrm{tr}\sigma\right)^{i},\nonumber 
\end{align}
\begin{align}
\left\langle \hat{n}^{*},\mathcal{H}\left(\hat{n},\,^{3}dx^{i}\right)\right\rangle  & =\,^{3}q^{ij}\left\langle \hat{n}^{*},\boldsymbol{\tau}_{j}+\boldsymbol{k}_{j}\right\rangle ,\label{eq:s2-1}\\
\left\langle \,^{3}dx^{i},\mathcal{H}\left(\hat{n},\,^{3}dx^{j}\right)\right\rangle  & =\,^{3}q^{jk}\left(\tau_{\;k}^{i}+k_{\;k}^{i}\right),\nonumber 
\end{align}
\begin{align}
\left\langle \hat{n}^{*},\mathcal{H}\left(\partial_{i},\hat{n}^{*}\right)\right\rangle  & =\left\langle \hat{n}^{*},\boldsymbol{\tau}_{i}-\boldsymbol{k}_{i}\right\rangle ,\label{eq:s3-1}\\
\left\langle \,^{3}dx^{i},\mathcal{H}\left(\partial_{j},\hat{n}^{*}\right)\right\rangle  & =\tau_{\;j}^{i}-k_{\;j}^{i},\nonumber 
\end{align}
\begin{align}
\left\langle \hat{n}^{*},\mathcal{H}\left(\partial_{i},\,^{3}dx^{j}\right)\right\rangle  & =-\frac{1}{4}\delta_{i}^{j}\mathcal{D}\left(\hat{n}\right)+\,^{3}q^{jk}\left(\left\langle \hat{n}^{*},\boldsymbol{\sigma}_{ik}+\varepsilon_{ikm}\boldsymbol{l}^{m}\right\rangle \right),\label{eq:s4-1}\\
\left\langle \,^{3}dx^{i},\mathcal{H}\left(\partial_{j},\,^{3}dx^{k}\right)\right\rangle  & =\frac{1}{4}\delta_{j}^{k}\mathcal{D}^{i}+\,^{3}q^{kl}\left(\mathcal{\sigma}_{\:\:jl}^{i}+\varepsilon_{jlm}l^{im}\right).\nonumber 
\end{align}

Another important relation is the decomposition of $\mathrm{tr}\mathcal{H}=\mathcal{H}\left(\partial_{\mu},dx^{\mu}\right)=\mathcal{H}_{\:\:\mu}^{\alpha\,\:\mu}\partial_{\alpha}$:
\begin{equation}
\mathrm{tr}\mathcal{H}=-\left\langle \hat{n}^{*},\mathcal{H}\left(\partial_{\mu},dx^{\mu}\right)\right\rangle \hat{n}+\left\langle \,^{3}dx^{i},\mathcal{H}\left(\partial_{\mu},dx^{\mu}\right)\right\rangle \partial_{i}.\label{eq:p4}
\end{equation}
On the other hand, using $dx^{0}=-\hat{n}^{*}N^{-1}$ and $dx^{i}=\hat{n}^{*}N^{i}N^{-1}+\,^{3}dx^{i}$,
(\ref{eq:p4}) could be written as:
\begin{equation}
\mathrm{tr}\mathcal{H}=-\mathcal{H}\left(\hat{n},\hat{n}^{*}\right)+\mathcal{H}\left(\partial_{i},\,^{3}dx^{i}\right).\label{eq:p5}
\end{equation}
Hence, using (\ref{eq:p5}) and (\ref{eq:pentings}):
\begin{align}
\left\langle \hat{n}^{*},\mathrm{tr}\mathcal{H}\right\rangle  & =-\mathcal{D}\left(\hat{n}\right)-\left\langle \hat{n}^{*},\textrm{tr}\boldsymbol{\sigma}\right\rangle +\left\langle \hat{n}^{*},\,^{3}\mathrm{tr}\Lambda\right\rangle =-\mathcal{D}\left(\hat{n}\right),\label{eq:s5}\\
\left\langle \,^{3}dx^{i},\mathrm{tr}\mathcal{H}\right\rangle  & =\mathcal{D}^{i}-\left\langle \,^{3}dx^{i},\textrm{tr}\boldsymbol{\sigma}\right\rangle +\left\langle \,^{3}dx^{i},\,^{3}\mathrm{tr}\Lambda\right\rangle =\mathcal{D}^{i},\nonumber 
\end{align}
with $\,^{3}\mathrm{tr}\Lambda=\,^{3}q^{ij}\Lambda\left(\partial_{i},\partial_{j}\right)=\textrm{tr}\boldsymbol{\sigma}$
and $\,^{3}\mathrm{tr}\Omega=\,^{3}q^{ij}\Omega\left(\partial_{i},\partial_{j}\right)=0$.
The quantities (\ref{eq:s1-1})-(\ref{eq:s4-1}) and (\ref{eq:s5})
will be used for the (3+1) decomposition of the hypermomentum equation
(\ref{eq:EoM2}).

\subsubsection*{The (3+1) Hypermomentum Equation in Adapted  Coordinate}

The hypermomentum equation (\ref{eq:EoM2}) could be split into 4
equations as follows:
\begin{align}
 & n^{\mu}n^{\lambda}T_{\mu\;\;\;\lambda}^{\:\:\alpha}\partial_{\alpha}-n_{\beta}\left(\nabla_{\hat{n}}g^{\alpha\beta}\right)\partial_{\alpha}+n_{\beta}\left(\nabla_{\nu}g^{\nu\beta}\right)\hat{n}=\kappa n^{\mu}n_{\beta}\mathcal{H}_{\:\:\mu}^{\alpha\,\:\beta}\partial_{\alpha}-\frac{2}{3}n^{\mu}n_{\beta}\kappa\mathcal{H}_{\:\:\;\;\,\sigma}^{\left[\alpha\right|\,\:\sigma}\delta_{\mu}^{\left|\beta\right]}\partial_{\alpha},\label{eq:gokil}\\
 & n^{\mu}g^{i\lambda}T_{\mu\;\;\;\lambda}^{\:\:\alpha}\partial_{\alpha}+\frac{1}{2}\left(g_{\sigma\lambda}\nabla_{\nu}g^{\sigma\lambda}\right)\left(n^{\nu}g^{\alpha i}\partial_{\alpha}-g^{\nu i}\hat{n}\right)-\left(\nabla_{\hat{n}}g^{\alpha i}\right)\partial_{\alpha}+\left(\nabla_{\nu}g^{\nu i}\right)\hat{n}=\kappa n^{\mu}\mathcal{H}_{\:\:\mu}^{\alpha\,\:i}\partial_{\alpha}-n^{\mu}\delta_{\beta}^{i}\frac{2}{3}\kappa\mathcal{H}_{\:\:\;\;\,\sigma}^{\left[\alpha\right|\,\:\sigma}\delta_{\mu}^{\left|\beta\right]}\partial_{\alpha},\nonumber \\
 & n^{\lambda}T_{i\;\;\;\lambda}^{\:\:\alpha}\partial_{\alpha}+\frac{1}{2}\left(g_{\sigma\lambda}\nabla_{\nu}g^{\sigma\lambda}\right)\left(\delta_{i}^{\:\nu}\hat{n}-n^{\nu}\partial_{i}\right)-n_{\beta}\left(\nabla_{i}g^{\alpha\beta}\right)\partial_{\alpha}+n_{\beta}\left(\nabla_{\nu}g^{\nu\beta}\right)\partial_{i}=\kappa n_{\beta}\mathcal{H}_{\:\:i}^{\alpha\,\:\beta}\partial_{\alpha}-\delta_{i}^{\mu}n_{\beta}\frac{2}{3}\kappa\mathcal{H}_{\:\:\;\;\,\sigma}^{\left[\alpha\right|\,\:\sigma}\delta_{i}^{\left|\beta\right]}\partial_{\alpha},\nonumber \\
 & g^{j\lambda}T_{i\;\;\;\lambda}^{\:\:\alpha}\partial_{\alpha}+\frac{1}{2}\left(g_{\sigma\lambda}\nabla_{\nu}g^{\sigma\lambda}\right)\left(\delta_{i}^{\:\nu}g^{\alpha j}\partial_{\alpha}-g^{\nu j}\partial_{i}\right)-\left(\nabla_{i}g^{\alpha j}\right)\partial_{\alpha}+\left(\nabla_{\nu}g^{\nu j}\right)\partial_{i}=\kappa\mathcal{H}_{\:\:i}^{\alpha\,\:j}\partial_{\alpha}-\delta_{i}^{\mu}\delta_{\beta}^{j}\frac{2}{3}\kappa\mathcal{H}_{\:\:\;\;\,\sigma}^{\left[\alpha\right|\,\:\sigma}\delta_{\mu}^{\left|\beta\right]}\partial_{\alpha},\nonumber 
\end{align}
by contracting the $\mu,\beta$ indices in (\ref{eq:EoM2}) with four
combination of $\hat{n}$ and $\partial_{i}$ \cite{nuaing}. Note
that by definition (\ref{eq:efh}), one needs to be careful that the
quantity $\mathcal{H}_{\:\:\mu}^{\alpha\,\:i}\partial_{\alpha}\neq\mathcal{H}\left(\partial_{\mu},\,^{3}dx^{i}\right)$
but:
\[
\mathcal{H}_{\:\:\mu}^{\alpha\,\:i}\partial_{\alpha}=\mathcal{H}\left(\partial_{\mu},dx^{i}\right)=N^{i}N^{-1}\mathcal{H}\left(\partial_{\mu},\hat{n}^{*}\right)+\mathcal{H}\left(\partial_{\mu},\,^{3}dx^{i}\right).
\]
Only in the Gauss (normal) coordinate, $\mathcal{H}\left(\partial_{\mu},dx^{i}\right)=\mathcal{H}\left(\partial_{\mu},\,^{3}dx^{i}\right)$.
Using the torsion decomposition (\ref{eq:p})-(\ref{eq:s}) and some
useful relations in (\ref{eq:h})-(\ref{eq:g}), (\ref{eq:gokil})
becomes:
\begin{align}
 & T\left(\hat{n},\hat{n}\right)-\left(\nabla_{\hat{n}}g^{*}\right)\left(\cdot,\hat{n}^{*}\right)+\Phi\left(\hat{n}\right)\hat{n}=\kappa\left(\mathcal{H}\left(\hat{n},\hat{n}^{*}\right)+\frac{1}{3}\left(\mathrm{tr}\mathcal{H}+\left\langle \hat{n}^{*},\mathrm{tr}\mathcal{H}\right\rangle \hat{n}\right)\right).\label{eq:gokil2}\\
 & T\left(\hat{n},g^{*}\left(dx^{i}\right)\right)-\left(\nabla_{\hat{n}}g^{*}\right)\left(dx^{i},\cdot\right)+\left(\Phi^{i}-\frac{1}{2}\Psi^{i}\right)\hat{n}+\frac{1}{2}\Psi\left(\hat{n}\right)g^{*}\left(dx^{i}\right)=\kappa\mathcal{H}\left(\hat{n},dx^{i}\right)-\kappa\frac{1}{3}\left(\left\langle dx^{i},\hat{n}\right\rangle \mathrm{tr}\mathcal{H}-\left\langle dx^{i},\mathrm{tr}\mathcal{H}\right\rangle \hat{n}\right),\nonumber \\
 & T\left(\partial_{i},\hat{n}\right)-\left(\nabla_{i}g^{*}\right)\left(\hat{n}^{*},\cdot\right)+\frac{1}{2}\Psi_{i}\hat{n}+\left(\Phi\left(\hat{n}\right)-\frac{1}{2}\Psi\left(\hat{n}\right)\right)\partial_{i}=\kappa\left(\mathcal{H}\left(\partial_{i},\hat{n}^{*}\right)+\frac{1}{3}\left\langle n^{*},\mathrm{tr}\mathcal{H}\right\rangle \partial_{i}\right)\nonumber \\
 & g^{j\lambda}T\left(\partial_{i},\partial_{\lambda}\right)-\left(\nabla_{i}g^{*}\right)\left(\cdot,dx^{j}\right)+\frac{1}{2}\Psi_{i}g^{*}\left(\cdot,dx^{j}\right)+\left(\Phi^{j}-\frac{1}{2}\Psi^{j}\right)\partial_{i}=\kappa\left(\mathcal{H}\left(\partial_{i},dx^{j}\right)-\frac{1}{3}\left(\delta_{i}^{j}\mathrm{tr}\mathcal{H}-\left\langle dx^{j},\mathrm{tr}\mathcal{H}\right\rangle \partial_{i}\right)\right)\nonumber 
\end{align}
with:
\begin{align*}
 & \Phi\left(\hat{n}\right)=-\left(\nabla_{\hat{n}}g^{*}\right)\left(\hat{n}^{*},\hat{n}^{*}\right)-N^{-1}N^{i}\left(\nabla_{i}g^{*}\right)\left(\hat{n}^{*},\hat{n}^{*}\right)+\left(\nabla_{i}g^{*}\right)\left(dx^{i},\hat{n}^{*}\right)\\
 & \Phi^{i}=-\left(\nabla_{\hat{n}}g^{*}\right)\left(\hat{n}^{*},dx^{i}\right)-N^{-1}N^{j}\left(\nabla_{j}g^{*}\right)\left(\hat{n}^{*},dx^{i}\right)+\left(\nabla_{j}g^{*}\right)\left(dx^{j},dx^{i}\right)\\
 & \Psi\left(\hat{n}\right)=\left(N^{-2}N^{j}N_{j}-1\right)\left(\nabla_{\hat{n}}g^{*}\right)\left(\hat{n}^{*},\hat{n}^{*}\right)-2N^{-1}N_{j}\left(\nabla_{\hat{n}}g^{*}\right)\left(\hat{n}^{*},dx^{j}\right)+\,^{3}q_{jk}\left(\nabla_{\hat{n}}g^{*}\right)\left(dx^{j},dx^{k}\right)\\
 & \Psi_{i}=\left(N^{-2}N^{j}N_{j}-1\right)\left(\nabla_{i}g^{*}\right)\left(\hat{n}^{*},\hat{n}^{*}\right)-2N^{-1}N_{j}\left(\nabla_{i}g^{*}\right)\left(\hat{n}^{*},dx^{j}\right)+\,^{3}q_{jk}\left(\nabla_{i}g^{*}\right)\left(dx^{j},dx^{k}\right)
\end{align*}
and $\Psi^{i}=\,^{3}q^{ij}\Psi_{j}$.

Applying the torsion and non-metricity decomposition in (\ref{eq:T00}),
(\ref{eq:T0i}), (\ref{eq:Tij}) and (\ref{eq:q1})-(\ref{eq:q4})
to the LHS of (\ref{eq:gokil2}), and applying (\ref{eq:s1-1})-(\ref{eq:s4-1})
and (\ref{eq:s5}) to the RHS of (\ref{eq:gokil2}), one could rewrite
the four equations in terms of the additional variables as follows:
\begin{align}
\left(\mathcal{K}_{i}^{\:\:i}-K_{i}^{\;i}\right)\hat{n}+\left(\Delta^{i}-\alpha^{i}\right)\partial_{i} & =-\kappa\left(\frac{1}{4}\mathcal{D}\left(\hat{n}\right)+\left\langle \hat{n}^{*},\textrm{tr}\boldsymbol{\sigma}\right\rangle \right)\hat{n}+\kappa\left(\left(\textrm{tr}\sigma\right)^{i}+\frac{1}{12}\mathcal{D}^{i}\right)\partial_{i}\label{eq:eq1}
\end{align}
\begin{align}
\left(\Delta^{i}-2\Theta^{i}-g\left(\partial^{i}\hat{n},\hat{n}\right)+\,^{3}\nabla_{j}\,^{3}q^{ji}-\frac{1}{2}\,^{3}q_{jk}\,^{3}\nabla^{i}\,^{3}q^{jk}-N^{-1}N^{i}\left(K_{j}^{\;j}-\mathcal{K}_{j}^{\:\:j}\right)\right)\hat{n}\label{eq:eq2}\\
+\left(\Delta^{ji}-\mathcal{K}^{ij}+g\left(\partial^{i}\hat{n},\,^{3}dx^{j}\right)+\left(\Theta\left(\hat{n}\right)+\frac{1}{2}\,^{3}q_{kl}\nabla_{\hat{n}}\,^{3}q^{kl}\right)\,^{3}q^{ij}-\nabla_{\hat{n}}\,^{3}q^{ij}+N^{-1}N^{i}\left(\Delta^{j}-\alpha^{j}\right)\right)\partial_{j} & =\mathrm{RHS1,}\nonumber 
\end{align}
\begin{align}
\left(\frac{1}{2}\,^{3}q_{jk}\,^{3}\nabla_{i}\,^{3}q^{jk}-\Delta_{i}+g\left(\partial_{i}\hat{n},\hat{n}\right)\right)\hat{n}\label{eq:eq3}\\
+\left(\mathcal{K}_{j}^{\:\:j}-K_{j}^{\;j}+\Theta\left(\hat{n}\right)-\frac{1}{2}\,^{3}q_{jk}\nabla_{\hat{n}}\,^{3}q^{jk}+\right)\partial_{i}+\left(K_{i}^{\;j}-\Delta_{\;\:i}^{j}-g\left(\partial_{i}\hat{n},\,^{3}dx^{j}\right)\right)\partial_{j} & =\mathrm{RHS2,}\nonumber 
\end{align}
\begin{align}
\left(\mathcal{K}_{i}^{\:\:j}-K_{\;i}^{j}+\left(\frac{1}{2}\,^{3}q_{kl}\,^{3}\nabla_{i}\,^{3}q^{kl}-\Delta_{i}+g\left(\partial_{i}\hat{n},\hat{n}\right)\right)N^{j}N^{-1}\right)\hat{n}+\,^{3}q^{jk}\,^{3}T\left(\partial_{i},\partial_{k}\right)\label{eq:eq4}\\
+\left(\Delta^{j}-\alpha^{j}+\,^{3}\nabla_{k}\,^{3}q^{jk}-\Theta^{j}-\frac{1}{2}\,^{3}q_{kl}\,^{3}\nabla^{j}\,^{3}q^{kl}+\left(\Theta\left(\hat{n}\right)-\frac{1}{2}\,^{3}q_{kl}\nabla_{\hat{n}}\,^{3}q^{kl}+\mathcal{K}_{k}^{\:\:k}-K_{k}^{\;k}\right)N^{-1}N^{j}\right)\partial_{i}\nonumber \\
+\left(\left(\Theta_{i}+\frac{1}{2}\,^{3}q_{lm}\,^{3}\nabla_{i}\,^{3}q^{lm}\right)\,^{3}q^{jk}-\,^{3}\nabla_{i}\,^{3}q^{jk}+N^{j}N^{-1}\left(K_{i}^{\;k}-\Delta_{\;\:i}^{k}-g\left(\partial_{i}\hat{n},\,^{3}dx^{k}\right)\right)\right)\partial_{k} & =\mathrm{RHS3,}\nonumber 
\end{align}
with:
\begin{align*}
\mathrm{RHS1}= & \kappa\left(\frac{1}{3}\mathcal{D}^{i}-\left\langle \hat{n}^{*},\boldsymbol{\tau}^{i}+\boldsymbol{k}^{i}\right\rangle -N^{i}N^{-1}\left(\frac{1}{4}\mathcal{D}\left(\hat{n}\right)+\left\langle \hat{n}^{*},\textrm{tr}\boldsymbol{\sigma}\right\rangle \right)\right)\hat{n}+\kappa\left(\tau^{ji}+k^{ji}+N^{i}N^{-1}\left(\frac{1}{12}\mathcal{D}^{j}+\left(\textrm{tr}\sigma\right)^{j}\right)\right)\partial_{j},\\
\mathrm{RHS2}= & \kappa\left(-\left\langle \hat{n}^{*},\boldsymbol{\tau}_{i}-\boldsymbol{k}_{i}\right\rangle \hat{n}-\frac{1}{3}\mathcal{D}\left(\hat{n}\right)\partial_{i}+\left(\tau_{\;i}^{j}-k_{\;i}^{j}\right)\partial_{j}\right),\\
\mathrm{RHS3}= & \kappa\left(-\frac{1}{12}\delta_{i}^{j}\mathcal{D}\left(\hat{n}\right)-\,^{3}q^{jk}\left\langle \hat{n}^{*},\boldsymbol{\sigma}_{ik}+\varepsilon_{ikm}\boldsymbol{l}^{m}\right\rangle -N^{j}N^{-1}\left\langle \hat{n}^{*},\boldsymbol{\tau}_{i}-\boldsymbol{k}_{i}\right\rangle \right)\hat{n}+\kappa\frac{1}{3}\left(\mathcal{D}^{j}-N^{-1}N^{j}\mathcal{D}\left(\hat{n}\right)\right)\partial_{i}\\
 & +\kappa\left(-\frac{1}{12}\delta_{i}^{j}\mathcal{D}^{k}+\,^{3}q^{jl}\left(\mathcal{\sigma}_{\:\:il}^{k}+\varepsilon_{ilm}l^{km}\right)+N^{j}N^{-1}\left(\tau_{\;i}^{k}-k_{\;i}^{k}\right)\right)\partial_{k}
\end{align*}
Contracting (\ref{eq:eq1})-(\ref{eq:eq4}) with $\hat{n}^{*}$ and
$\,^{3}dx^{i}$, we obtain 8 equations. (\ref{eq:eq1}) gives:
\begin{align*}
 & \mathcal{K}_{i}^{\:\:i}-K_{i}^{\;i}=-\kappa\left(\frac{1}{4}\mathcal{D}\left(\hat{n}\right)+\left\langle \hat{n}^{*},\textrm{tr}\boldsymbol{\sigma}\right\rangle \right),\\
 & \Delta^{i}-\alpha^{i}=\kappa\left(\left(\textrm{tr}\sigma\right)^{i}+\frac{1}{12}\mathcal{D}^{i}\right);
\end{align*}
(\ref{eq:eq2}) gives:
\begin{align*}
\Delta^{i}-2\Theta^{i}-g\left(\partial^{i}\hat{n},\hat{n}\right)+\,^{3}\nabla_{j}\,^{3}q^{ji}-\frac{1}{2}\,^{3}q_{jk}\,^{3}\nabla^{i}\,^{3}q^{jk}-N^{-1}N^{i}\left(K_{j}^{\;j}-\mathcal{K}_{j}^{\:\:j}\right)= & \kappa\left(\frac{1}{3}\mathcal{D}^{i}-\left\langle \hat{n}^{*},\boldsymbol{\tau}^{i}+\boldsymbol{k}^{i}\right\rangle \right)\\
 & -\kappa N^{i}N^{-1}\left(\frac{1}{4}\mathcal{D}\left(\hat{n}\right)+\left\langle \hat{n}^{*},\textrm{tr}\boldsymbol{\sigma}\right\rangle \right),
\end{align*}
\begin{align*}
\Delta^{ji}-\mathcal{K}^{ij}+g\left(\partial^{i}\hat{n},\,^{3}dx^{j}\right)-\nabla_{\hat{n}}\,^{3}q^{ij}+N^{-1}N^{i}\left(\Delta^{j}-\alpha^{j}\right)\\
+\left(\Theta\left(\hat{n}\right)+\frac{1}{2}\,^{3}q_{kl}\nabla_{\hat{n}}\,^{3}q^{kl}\right)\,^{3}q^{ij} & =\kappa\left(\tau^{ji}+k^{ji}+N^{i}N^{-1}\left(\frac{1}{12}\mathcal{D}^{j}+\left(\textrm{tr}\sigma\right)^{j}\right)\right).
\end{align*}
(\ref{eq:eq3}) gives:
\begin{align*}
 & \frac{1}{2}\,^{3}q_{jk}\,^{3}\nabla_{i}\,^{3}q^{jk}-\Delta_{i}+g\left(\partial_{i}\hat{n},\hat{n}\right)=-\kappa\left\langle \hat{n}^{*},\boldsymbol{\tau}_{i}-\boldsymbol{k}_{i}\right\rangle ,\\
 & \left(\mathcal{K}_{l}^{\:\:l}-K_{l}^{\;l}+\Theta\left(\hat{n}\right)-\frac{1}{2}\,^{3}q_{kl}\nabla_{\hat{n}}\,^{3}q^{kl}+\right)\delta_{i}^{j}+K_{i}^{\;j}-\Delta_{\;\:i}^{j}-g\left(\partial_{i}\hat{n},\,^{3}dx^{j}\right)=\kappa\left(-\frac{1}{3}\mathcal{D}\left(\hat{n}\right)\delta_{i}^{j}+\tau_{\;i}^{j}-k_{\;i}^{j}\right),
\end{align*}
and finally, (\ref{eq:eq4}) gives:
\begin{align*}
\mathcal{K}_{i}^{\:\:j}-K_{\;i}^{j}+\left(\frac{1}{2}\,^{3}q_{kl}\,^{3}\nabla_{i}\,^{3}q^{kl}-\Delta_{i}+g\left(\partial_{i}\hat{n},\hat{n}\right)\right)N^{j}N^{-1}= & \kappa\left(-\frac{1}{12}\delta_{i}^{j}\mathcal{D}\left(\hat{n}\right)-\,^{3}q^{jk}\left\langle \hat{n}^{*},\boldsymbol{\sigma}_{ik}+\varepsilon_{ikm}\boldsymbol{l}^{m}\right\rangle \right)\\
 & -\kappa N^{j}N^{-1}\left\langle \hat{n}^{*},\boldsymbol{\tau}_{i}-\boldsymbol{k}_{i}\right\rangle ,
\end{align*}
\begin{align*}
\,^{3}T_{i}^{\,\,kj}+\left(\Theta_{i}+\frac{1}{2}\,^{3}q_{lm}\,^{3}\nabla_{i}\,^{3}q^{lm}\right)\,^{3}q^{jk}-\,^{3}\nabla_{i}\,^{3}q^{jk}+N^{j}N^{-1}\left(K_{i}^{\;k}-\Delta_{\;\:i}^{k}-g\left(\partial_{i}\hat{n},\,^{3}dx^{k}\right)\right)\\
+\left(\Delta^{j}-\alpha^{j}+\,^{3}\nabla_{l}\,^{3}q^{jl}-\Theta^{j}-\frac{1}{2}\,^{3}q_{lm}\,^{3}\nabla^{j}\,^{3}q^{lm}+\left(\Theta\left(\hat{n}\right)-\frac{1}{2}\,^{3}q_{lm}\nabla_{n}\,^{3}q^{lm}+\mathcal{K}_{l}^{\:\:l}-K_{l}^{\;l}\right)N^{-1}N^{j}\right)\delta_{i}^{k} & =\mathrm{RHS4,}
\end{align*}
with:
\begin{equation}
\mathrm{RHS4}=\kappa\left(\frac{1}{3}\left(\mathcal{D}^{j}-N^{-1}N^{j}\mathcal{D}\left(\hat{n}\right)\right)\delta_{i}^{k}-\frac{1}{12}\delta_{i}^{j}\mathcal{D}^{k}+\,^{3}q^{jl}\left(\mathcal{\sigma}_{\:\:il}^{k}+\varepsilon_{ilm}l^{km}\right)+N^{j}N^{-1}\left(\tau_{\;i}^{k}-k_{\;i}^{k}\right)\right).\label{eq:rhs4}
\end{equation}

\subsubsection*{The Traceless Torsion Constraint Decomposition}

The hypermomentum equation (\ref{eq:EoM2}) provides 60 independent
equations, 4 more equations are needed to determine uniquely the connection
from the equation of motions (\ref{eq:EoM1})-(\ref{eq:EoM2}). These
vectorial degrees of freedom are reduced by traceless torsion constraint
(\ref{eq:EoM3}), consult \cite{Iosifidis,Iosifidis4} or \cite{nuaing}
for some detailed explanations. Let us write $\boldsymbol{T}=T_{\nu}dx^{\nu}\in T_{p}^{*}\mathcal{M}$
as a 1-form where its components satisfy:
\[
T_{\nu}=T_{\mu\;\nu}^{\;\mu}=\left\langle dx^{\mu},T\left(\partial_{\mu},\partial_{\nu}\right)\right\rangle =0.
\]
Contracting $\boldsymbol{T}$ with the normal $\hat{n}$ and $\partial_{i}$,
and using $dx^{0}=-\hat{n}^{*}N^{-1}$ and $dx^{i}=\hat{n}^{*}N^{i}N^{-1}+\,^{3}dx^{i}$
gives:
\begin{align*}
 & n^{\nu}T_{\nu}=n^{\nu}T_{\mu\;\nu}^{\;\mu}=-\left\langle \,^{3}dx^{i},T\left(\partial_{i},\hat{n}\right)\right\rangle ,\\
 & T_{i}=T_{\mu\;i}^{\;\mu}=-\left\langle \hat{n}^{*},T\left(\hat{n},\partial_{i}\right)\right\rangle +\left\langle \,^{3}dx^{j},T\left(\partial_{j},\partial_{i}\right)\right\rangle ,
\end{align*}
then using torsion decomposition (\ref{eq:T00})-(\ref{eq:T0i})-(\ref{eq:Tij}),
one obtains \cite{nuaing}:
\begin{align}
 & \left\langle \boldsymbol{T},\hat{n}\right\rangle =n^{\nu}T_{\nu}=n^{\nu}T_{\mu\;\nu}^{\;\mu}=\Delta_{\;\:i}^{i}-\mathcal{K}_{i}^{\:\:i}+\left\langle \,^{3}dx^{i},\partial_{i}\hat{n}\right\rangle =0,\label{eq:proj1}\\
 & \left\langle \boldsymbol{T},\partial_{i}\right\rangle =T_{i}=T_{\mu\;i}^{\;\mu}=\Delta_{i}-\Theta_{i}-\left\langle \hat{n}^{*},\partial_{i}\hat{n}\right\rangle +\,^{3}T_{j\;i}^{\;j}=0.\label{eq:proj2}
\end{align}

\subsubsection*{Classification of the Equations}

The hypermomentum equation (\ref{eq:EoM2}), together with the traceless
torsion constraint (\ref{eq:EoM3}), provide 10 equations. Collecting
and classifying all these results, we have 2 scalar equations, resulting
from the hypermomentum equation and the traceless torsion equation:
\begin{align*}
 & \mathcal{K}_{i}^{\:\:i}-K_{i}^{\;i}=-\kappa\left(\frac{1}{4}\mathcal{D}\left(\hat{n}\right)+\left\langle \hat{n}^{*},\textrm{tr}\boldsymbol{\sigma}\right\rangle \right),\\
 & \Delta_{\;\:i}^{i}-\mathcal{K}_{i}^{\:\:i}+g\left(\partial_{i}\hat{n},\,^{3}dx^{i}\right)=0;
\end{align*}
4 vectors equations, 3 from the hypermomentum equations:
\begin{align*}
\Delta^{i}-\alpha^{i} & =\kappa\left(\left(\textrm{tr}\sigma\right)^{i}+\frac{1}{12}\mathcal{D}^{i}\right),\\
\frac{1}{2}\,^{3}q_{jk}\,^{3}\nabla_{i}\,^{3}q^{jk}-\Delta_{i}+g\left(\partial_{i}\hat{n},\hat{n}\right) & =\kappa\left\langle \hat{n}^{*},\boldsymbol{k}_{i}-\boldsymbol{\tau}_{i}\right\rangle ,
\end{align*}
\begin{align*}
\Delta^{i}-2\Theta^{i}-g\left(\partial^{i}\hat{n},\hat{n}\right)+\,^{3}\nabla_{j}\,^{3}q^{ji}-\frac{1}{2}\,^{3}q_{jk}\,^{3}\nabla^{i}\,^{3}q^{jk}-N^{-1}N^{i}\left(K_{j}^{\;j}-\mathcal{K}_{j}^{\:\:j}\right)= & \kappa\left(\frac{1}{3}\mathcal{D}^{i}-\left\langle \hat{n}^{*},\boldsymbol{\tau}^{i}+\boldsymbol{k}^{i}\right\rangle \right)\\
 & -\kappa N^{i}N^{-1}\left(\frac{1}{4}\mathcal{D}\left(\hat{n}\right)+\left\langle \hat{n}^{*},\textrm{tr}\boldsymbol{\sigma}\right\rangle \right),
\end{align*}
and 1 from the traceless torsion constraint:
\[
\Delta_{i}-\Theta_{i}-\left\langle \hat{n}^{*},\partial_{i}\hat{n}\right\rangle +\,^{3}T_{j\;i}^{\,\;j}=0;
\]
3 matrix equations:
\begin{align*}
\Delta^{ji}-\mathcal{K}^{ij}+g^{*}\left(\partial^{i}\hat{n},\,^{3}dx^{j}\right)-\nabla_{\hat{n}}\,^{3}q^{ij}+N^{-1}N^{i}\left(\Delta^{j}-\alpha^{j}\right)\\
+\left(\Theta\left(\hat{n}\right)+\frac{1}{2}\,^{3}q_{kl}\nabla_{\hat{n}}\,^{3}q^{kl}\right)\,^{3}q^{ij} & =\kappa\left(\tau^{ji}+k^{ji}+N^{i}N^{-1}\left(\frac{1}{12}\mathcal{D}^{j}+\left(\textrm{tr}\sigma\right)^{j}\right)\right),
\end{align*}
\begin{align*}
\left(\mathcal{K}_{l}^{\:\:l}-K_{l}^{\;l}+\Theta\left(\hat{n}\right)-\frac{1}{2}\,^{3}q_{kl}\nabla_{\hat{n}}\,^{3}q^{kl}+\right)\delta_{i}^{j}+K_{i}^{\;j}-\Delta_{\;\:i}^{j}-\left\langle \,^{3}dx^{j},\partial_{i}\hat{n}\right\rangle =\kappa\left(-\frac{1}{3}\mathcal{D}\left(\hat{n}\right)\delta_{i}^{j}+\tau_{\;i}^{j}-k_{\;i}^{j}\right),
\end{align*}
\begin{align*}
\mathcal{K}_{i}^{\:\:j}-K_{\;i}^{j}+\left(\frac{1}{2}\,^{3}q_{kl}\,^{3}\nabla_{i}\,^{3}q^{kl}-\Delta_{i}+g\left(\partial_{i}\hat{n},\hat{n}\right)\right)N^{j}N^{-1}= & \kappa\left(-\frac{1}{12}\delta_{i}^{j}\mathcal{D}\left(\hat{n}\right)-\,^{3}q^{jk}\left\langle \hat{n}^{*},\boldsymbol{\sigma}_{ik}+\varepsilon_{ikm}\boldsymbol{l}^{m}\right\rangle \right)\\
 & -\kappa\left(N^{j}N^{-1}\left\langle \hat{n}^{*},\boldsymbol{\tau}_{i}-\boldsymbol{k}_{i}\right\rangle \right),
\end{align*}
and finally 1 tensor equation of order-$\tbinom{2}{1}$:
\begin{align*}
\,^{3}T_{i}^{\,\,kj}+\left(\left(\Theta_{i}+\frac{1}{2}\,^{3}q_{lm}\,^{3}\nabla_{i}\,^{3}q^{lm}\right)\,^{3}q^{jk}-\,^{3}\nabla_{i}\,^{3}q^{jk}+N^{j}N^{-1}\left(K_{i}^{\;k}-\Delta_{\;\:i}^{k}-g\left(\partial_{i}\hat{n},\,^{3}dx^{k}\right)\right)\right)\\
+\left(\Delta^{j}-\alpha^{j}+\,^{3}\nabla_{l}\,^{3}q^{jl}-\Theta^{j}-\frac{1}{2}\,^{3}q_{lm}\,^{3}\nabla^{j}\,^{3}q^{lm}+\left(\Theta\left(\hat{n}\right)-\frac{1}{2}\,^{3}q_{lm}\nabla_{\hat{n}}\,^{3}q^{lm}+\mathcal{K}_{l}^{\:\:l}-K_{l}^{\;l}\right)N^{-1}N^{j}\right)\delta_{i}^{k} & =\mathrm{RHS4,}
\end{align*}
with RHS4 satisfies (\ref{eq:rhs4}). These 10 equations are the main
result in this article, that are, the complete (3+1) decomposition
of (\ref{eq:EoM2}) and (\ref{eq:EoM3}) in the adapted (not necessarily
normal) coordinate.

\section{Special Cases}

\subsubsection*{Special Case I: Zero Hypermomentum}

Let us take a special case where the hypermomentum $\mathcal{H}$
vanishes. The scalar equations are simplified as follows:
\begin{align}
 & \mathcal{K}_{i}^{\:\:i}-K_{i}^{\;i}=0,\label{eq:sclr-1-1}\\
 & \Delta_{\;\:i}^{i}-\mathcal{K}_{i}^{\:\:i}+\left\langle \,^{3}dx^{i},\partial_{i}\hat{n}\right\rangle =0,\label{eq:scla}
\end{align}
With the scalar equation (\ref{eq:sclr-1-1}), the vector equations
becomes:
\begin{align}
 & \Delta^{i}-\alpha^{i}=0,\label{eq:v1}\\
 & \Theta^{i}=\frac{1}{2}\,^{3}\nabla_{j}\,^{3}q^{ji}.\label{eq:v2}
\end{align}
\begin{align}
\Delta^{i}-\Theta^{i}-g\left(\partial^{i}\hat{n},\hat{n}\right) & =\frac{1}{2}\left(\,^{3}q_{jk}\,^{3}\nabla^{i}\,^{3}q^{jk}-\,^{3}\nabla_{j}\,^{3}q^{ji}\right).\label{eq:v3}\\
\Delta_{i}-\Theta_{i}-g\left(\partial_{i}\hat{n},\hat{n}\right) & =-\,^{3}T_{j\;i}^{\;j}.\label{eq:v4}
\end{align}
Inserting (\ref{eq:sclr-1-1}) and using vector equations (\ref{eq:v2})-(\ref{eq:v3})
together with the projective invariant constraint (\ref{eq:v4}),
the matrix equations becomes:
\begin{align}
\left(\Theta\left(\hat{n}\right)+\frac{1}{2}\,^{3}q_{kl}\left(\hat{n}\left[\,^{3}q^{kl}\right]+\Delta^{kl}+\Delta^{lk}\right)\right)\,^{3}q^{ij}-\hat{n}\left[\,^{3}q^{ij}\right]-\Delta^{ij}-\mathcal{K}^{ij}+\left\langle \,^{3}dx^{j},\partial^{i}\hat{n}\right\rangle =0,\label{eq:m1}\\
\left(\Theta\left(\hat{n}\right)-\frac{1}{2}\,^{3}q_{kl}\left(\hat{n}\left[\,^{3}q^{kl}\right]+\Delta^{kl}+\Delta^{lk}\right)\right)\,^{3}q^{ij}+K^{ij}-\Delta^{ji}-\left\langle \,^{3}dx^{j},\partial^{i}\hat{n}\right\rangle =0,\label{eq:m2}\\
\mathcal{K}^{ij}-K^{ji}=0.\label{eq:m3}
\end{align}
Contracting (\ref{eq:m1}) and (\ref{eq:m2}) with $q_{ij}$, and
then using both the scalar equations, gives:
\begin{align*}
3\Theta\left(\hat{n}\right)=-\frac{1}{2}\,^{3}q_{kl}\left(\hat{n}\left[\,^{3}q^{kl}\right]+\Delta^{kl}+\Delta^{lk}\right),\\
3\Theta\left(\hat{n}\right)=\frac{3}{2}\,^{3}q_{kl}\left(\hat{n}\left[\,^{3}q^{kl}\right]+\Delta^{kl}+\Delta^{lk}\right),
\end{align*}
which is only satisfied if:
\begin{align}
 & \Theta\left(\hat{n}\right)=0,\label{eq:m1a}\\
 & \hat{n}\left[\,^{3}q^{kl}\right]+\Delta^{kl}+\Delta^{lk}=0.\label{eq:m2a}
\end{align}
Inserting (\ref{eq:m1a}) and (\ref{eq:m2a}) to (\ref{eq:m1})-(\ref{eq:m3})
gives:
\begin{align}
\Delta^{ji}-\mathcal{K}^{ij}+\left\langle \,^{3}dx^{j},\partial^{i}\hat{n}\right\rangle =0,\label{eq:m1b}\\
K^{ij}-\mathcal{K}^{ij}=0,\label{eq:m2b}\\
K^{ij}-K^{ji}=0.\label{eq:m3b}
\end{align}

The last one is the tensor equation. Raising all the indices, using
vector equations (\ref{eq:v1})-(\ref{eq:v4}) and matrix equations
(\ref{eq:m1b})-(\ref{eq:m3b}) gives:
\begin{align}
\frac{1}{2}\left(\,^{3}\nabla_{l}\,^{3}q^{li}+\,^{3}q_{lm}\,^{3}\nabla^{i}\,^{3}q^{lm}\right)\,^{3}q^{jk}-\,^{3}\nabla^{i}\,^{3}q^{jk}+N^{j}N^{-1}\left(-\Delta^{ki}+\mathcal{K}^{ik}-\left\langle \,^{3}dx^{k},\partial^{i}\hat{n}\right\rangle \right)\label{eq:tensor}\\
+\frac{1}{2}\left(\,^{3}\nabla_{l}\,^{3}q^{lj}-\,^{3}q_{lm}\,^{3}\nabla^{j}\,^{3}q^{lm}\right)\,^{3}q^{ik}+\,^{3}T^{ikj} & =0.\nonumber 
\end{align}
Contracting (\ref{eq:tensor}) with $q_{jk}$ gives:
\begin{align*}
\,^{3}\nabla_{j}\,^{3}q^{ij}-\,^{3}q_{jk}\,^{3}\nabla^{i}\,^{3}q^{jk}-q_{jk}\,^{3}T^{ikj} & =0,
\end{align*}
then decomposing this tensor equation into the symmetric and antisymmetric
parts of the $\left(i,j\right)$-indices, gives:
\begin{equation}
\,^{3}\nabla_{j}\,^{3}q^{\left(ij\right)}-\,^{3}q_{jk}\,^{3}\nabla^{\left(i\right.}\,^{3}q^{\left.j\right)k}=0,\label{eq:tensorsym}
\end{equation}
and 
\begin{equation}
\,^{3}T_{\;j}^{j\:\:i}-\,^{3}q_{ik}\,^{3}\nabla^{\left[i\right.}\,^{3}q^{\left.j\right]k}=0.\label{eq:tensorantsym}
\end{equation}

Let us solve the simplified equations (\ref{eq:sclr-1-1})-(\ref{eq:scla}),
(\ref{eq:v1})-(\ref{eq:v4}), (\ref{eq:m1b})-(\ref{eq:m3b}), and
(\ref{eq:tensorsym})-(\ref{eq:tensorantsym}). The easiest way is
to start from (\ref{eq:tensorsym}), which is satisfied if:
\begin{equation}
\,^{3}\nabla^{i}\,^{3}q^{jk}=0.\label{eq:1}
\end{equation}
Inserting (\ref{eq:1}) to (\ref{eq:tensor}) gives:
\begin{align}
\,^{3}T^{ikj} & =N^{j}N^{-1}\left(\Delta^{ki}-\mathcal{K}^{ik}+\left\langle \,^{3}dx^{k},\partial^{i}\hat{n}\right\rangle \right).\label{eq:2}
\end{align}
Now we have the following equations (\ref{eq:sclr-1-1})-(\ref{eq:scla}),
(\ref{eq:v1})-(\ref{eq:v4}), (\ref{eq:m1a})-(\ref{eq:m2a}), (\ref{eq:m1b})-(\ref{eq:m3b}),
and (\ref{eq:1})-(\ref{eq:2}). Solving these equations gives the
torsionless condition:
\begin{equation}
\begin{array}{ccc}
K^{ij}-K^{ji}=0, & \qquad & \Delta_{i}-\Theta_{i}-\left\langle \hat{n}^{*},\partial_{i}\hat{n}\right\rangle =0,\\
 & \,\\
\,^{3}T^{ikj}=0, &  & \Delta^{ji}-\mathcal{K}^{ij}+\left\langle \,^{3}dx^{j},\partial^{i}\hat{n}\right\rangle =0,
\end{array}\label{eq:torsionless}
\end{equation}
and metric compatibility:
\begin{equation}
\begin{array}{ccc}
\Theta\left(\hat{n}\right)=0, & \qquad & K^{ij}-\mathcal{K}^{ij}=0,\\
 & \,\\
\Theta^{i}=0, &  & \,^{3}\nabla_{\hat{n}}\,^{3}q^{jk},=0\\
 & \,\\
\Delta^{i}-\alpha^{i}=0, &  & \,^{3}\nabla_{i}\,^{3}q^{jk},=0
\end{array}\label{eq:metricomp}
\end{equation}

We could  conclude that for zero hypermomentum, the affine connection
becomes Levi-Civita, and the (3+1) stress-energy-momentum equations
for MAGR return to the original EFE, see \cite{nuaing}. Furthermore,
without the traceless torsion constraint (\ref{eq:proj1})-(\ref{eq:proj2}),
it is not possible to retrieve the metric compatibility and torsionless
condition in the absence of hypermomentum \cite{nuaing}.

\subsubsection*{Special Case II: Metric Connection in Normal Coordinate}

For the next two subsections, we will use a special coordinate where
the lapse $N=1$ and the shift $\boldsymbol{N}=0$. The choice of
coordinates will not alter the physical interpretation, but will greatly
simplifies our calculation. For the metric connection (not necessarily
torsionless), we apply the metricity compatibility (\ref{eq:metricomp})
to the 8 (3+1) equations in the normal coordinate. Notice that the
metricity condition causes the Weyl tensor $Q_{\mu}=0$ (but not vice
versa), and this guarantees the consistency of the equations of motion
under the projective invariance transformation \cite{Iosifidis4}.
The connection could be uniquely determined only from (\ref{eq:EoM1})
and (\ref{eq:EoM2}), hence, one does not need to consider the traceless
torsion constraint (\ref{eq:EoM3}).

With the metricity condition (\ref{eq:metricomp}), the 8 additional
variables reduces into 4, where we choose to work with only $\Delta^{i},$
$K^{ij},$ $\Delta^{ij}$, and $\,^{3}T_{i\;\:\;k}^{\:\:j}$. The
(3+1) equations consist of 1 scalar equation:
\begin{align}
-\kappa\left(\frac{1}{4}\mathcal{D}\left(\hat{n}\right)+\left\langle \hat{n}^{*},\textrm{tr}\boldsymbol{\sigma}\right\rangle \right)\equiv0,\label{eq:con1}
\end{align}
3 vector equations:
\begin{align}
 & \kappa\left(\left(\textrm{tr}\sigma\right)^{i}+\frac{1}{12}\mathcal{D}^{i}\right)\equiv0,\label{eq:con2}\\
 & \Delta^{i}=\kappa\left(\frac{1}{3}\mathcal{D}^{i}-\left\langle \hat{n}^{*},\boldsymbol{\tau}^{i}+\boldsymbol{k}^{i}\right\rangle \right),\nonumber \\
 & \Delta^{i}=-\kappa\left\langle \hat{n}^{*},\boldsymbol{k}^{i}-\boldsymbol{\tau}^{i}\right\rangle ,\nonumber 
\end{align}
3 matrix equations:
\begin{align*}
 & \Delta^{ji}-K^{ij}=\kappa\left(\tau^{ji}+k^{ji}\right),\\
 & K_{i}^{\;j}-\Delta_{\;\:i}^{j}=\kappa\left(-\frac{1}{3}\mathcal{D}\left(\hat{n}\right)\delta_{i}^{j}+\tau_{\;i}^{j}-k_{\;i}^{j}\right),\\
 & K_{i}^{\:\:j}-K_{\;i}^{j}=\kappa\left(-\frac{1}{12}\delta_{i}^{j}\mathcal{D}\left(\hat{n}\right)-\,^{3}q^{jk}\left\langle \hat{n}^{*},\boldsymbol{\sigma}_{ik}+\varepsilon_{ikm}\boldsymbol{l}^{m}\right\rangle \right),
\end{align*}
and 1 tensor equation:
\begin{align*}
\,^{3}T_{i}^{\,\,kj} & =\kappa\left(\frac{1}{3}\left(\mathcal{D}^{j}\right)\delta_{i}^{k}-\frac{1}{12}\delta_{i}^{j}\mathcal{D}^{k}+\,^{3}q^{jl}\left(\mathcal{\sigma}_{\:\:il}^{k}+\varepsilon_{ilm}l^{km}\right)\right).
\end{align*}

The scalar equation and the first vector equation are constraints
on the hypermomentum $\mathcal{H}$. The remaining two vector equations,
which could be written compactly as:
\begin{align}
\Delta^{i}= & \kappa\left(\frac{1}{3}\mathcal{D}^{i}-\left\langle \hat{n}^{*},\boldsymbol{\tau}^{i}+\boldsymbol{k}^{i}\right\rangle \right),\nonumber \\
\Delta^{i}= & \kappa\left\langle \hat{n}^{*},\boldsymbol{\tau}^{i}-\boldsymbol{k}^{i}\right\rangle ,\label{eq:deltai}
\end{align}
could be solved to obtain the additional variable $\Delta_{i}$ in
terms of hypermomentum in (\ref{eq:deltai}) and 1 constraint on $\mathcal{H}$:
\begin{align}
\frac{1}{3}\mathcal{D}^{i}-2\left\langle \hat{n}^{*},\boldsymbol{\tau}^{i}\right\rangle =0.\label{eq:con3}
\end{align}

The matrix equations could be compactly written as follows:
\begin{align}
K^{ij}-\Delta^{ji} & =-\kappa\left(\tau^{ji}+k^{ji}\right),\label{eq:met1}\\
K^{ij}-\Delta^{ji} & =\kappa\left(-\frac{1}{3}\mathcal{D}\left(\hat{n}\right)\,^{3}q^{ij}+\tau^{ji}-k^{ji}\right),\label{eq:met2}\\
K^{ij}-K^{ji} & =\kappa\left(-\frac{1}{12}\,^{3}q^{ij}\mathcal{D}\left(\hat{n}\right)-\,^{3}q^{in}\,^{3}q^{jk}\left\langle \hat{n}^{*},\boldsymbol{\sigma}_{nk}+\varepsilon_{nkm}\boldsymbol{l}^{m}\right\rangle \right).\nonumber 
\end{align}
Solving the first and second gives another constraint on $\mathcal{H}$:
\begin{align}
2\tau^{ji}-\frac{1}{3}\mathcal{D}\left(\hat{n}\right)\,^{3}q^{ij} & =0.\label{eq:con4}
\end{align}
Now let us focus on the last matrix equation, which could be written
as:
\begin{align*}
2K^{\left[ij\right]}= & \kappa\left(-\frac{1}{12}\,^{3}q^{ij}\mathcal{D}\left(\hat{n}\right)-\left\langle \hat{n}^{*},\boldsymbol{\sigma}^{\left(ij\right)}\right\rangle -\left\langle \hat{n}^{*},\varepsilon_{\;\:m}^{ij}\boldsymbol{l}^{m}\right\rangle \right),
\end{align*}
 notice that $\boldsymbol{\sigma}^{\left[ij\right]}=0$ since $\boldsymbol{\sigma}^{ij}$
is symmetric, and:
\[
\,^{3}q^{in}\,^{3}q^{jk}\left\langle \hat{n}^{*},\varepsilon_{nkm}\boldsymbol{l}^{m}\right\rangle =\,^{3}q^{n\left[i\right|}\,^{3}q^{k\left|j\right]}\left\langle \hat{n}^{*},\varepsilon_{nkm}\boldsymbol{l}^{m}\right\rangle =\left\langle \hat{n}^{*},\varepsilon_{\;\:m}^{ij}\boldsymbol{l}^{m}\right\rangle 
\]
We could split this equation in its symmetric and antisymmetric part:
\begin{align}
\kappa\left(-\frac{1}{12}\,^{3}q^{ij}\mathcal{D}\left(\hat{n}\right)-\left\langle \hat{n}^{*},\boldsymbol{\sigma}^{\left(ij\right)}\right\rangle \right) & =0,\label{eq:con5}\\
-\frac{1}{2}\kappa\left\langle \hat{n}^{*},\varepsilon_{\;\:m}^{ij}\boldsymbol{l}^{m}\right\rangle  & =K^{\left[ij\right]}.\label{eq:antisym1}
\end{align}
(\ref{eq:con5}) is another constraint on $\mathcal{H}.$

Let us solve (\ref{eq:met1})-(\ref{eq:met2}), that could be written
as:
\begin{align*}
 & K^{ij}+\hat{n}\left[\,^{3}q^{ij}\right]+\Delta^{ij}=-\kappa\left(\tau^{ji}+k^{ji}\right),\\
 & K^{ij}-\Delta^{ji}=\kappa\left(-\frac{1}{3}\mathcal{D}\left(\hat{n}\right)\,^{3}q^{ij}+\tau^{ji}-k^{ji}\right),
\end{align*}
using (\ref{eq:q3}) and the fact that $\,^{3}\nabla_{\hat{n}}\,^{3}q^{*}=0$
for metric connection. Subtituting and eliminating these two equations
give:
\begin{align*}
2K^{ij}+\hat{n}\left[\,^{3}q^{ij}\right]+2\Delta^{\left[ij\right]} & =-\kappa\left(2k^{ji}+\frac{1}{3}\mathcal{D}\left(\hat{n}\right)\,^{3}q^{ij}\right),\\
\hat{n}\left[\,^{3}q^{ij}\right]+2\Delta^{\left(ij\right)} & =-\kappa\left(2\tau^{ji}-\frac{1}{3}\mathcal{D}\left(\hat{n}\right)\,^{3}q^{ij}\right).
\end{align*}
If we split these two into their symmetric and antisymmetric parts,
we have:
\begin{align}
K^{\left(ij\right)} & =-\kappa\left(k^{\left(ji\right)}+\frac{1}{6}\mathcal{D}\left(\hat{n}\right)\,^{3}q^{ij}\right)-\frac{1}{2}\hat{n}\left[\,^{3}q^{ij}\right].\label{eq:a1}\\
K^{\left[ij\right]}+\Delta^{\left[ij\right]} & =-\kappa k^{\left[ji\right]}.\label{eq:a2}
\end{align}
\begin{align}
\Delta^{\left(ij\right)} & =-\kappa\left(\tau^{\left(ji\right)}-\frac{1}{6}\mathcal{D}\left(\hat{n}\right)\,^{3}q^{ij}\right)-\frac{1}{2}\hat{n}\left[\,^{3}q^{ij}\right].\label{eq:a3}\\
-\kappa\tau^{\left[ji\right]} & =0.\label{eq:con6}
\end{align}
(\ref{eq:con6}) is another constraint on $\mathcal{H}.$ Using (\ref{eq:antisym1}),
(\ref{eq:a1}), (\ref{eq:a2}), and (\ref{eq:a3}), we could solve
for $K^{\left(ij\right)}$, $K^{\left[ij\right]}$, $\Delta^{\left(ij\right)}$,
and $\Delta^{\left[ij\right]}$ to obtain:
\begin{align*}
K^{ij} & =\kappa\left(-\frac{1}{6}\mathcal{D}\left(\hat{n}\right)\,^{3}q^{ij}-\frac{1}{2}\left\langle \hat{n}^{*},\varepsilon_{\;\:m}^{ij}\boldsymbol{l}^{m}\right\rangle -k^{\left(ji\right)}\right)-\frac{1}{2}\hat{n}\left[\,^{3}q^{ij}\right],\\
\Delta^{ij} & =\kappa\left(\frac{1}{6}\mathcal{D}\left(\hat{n}\right)\,^{3}q^{ij}+\frac{1}{2}\left\langle \hat{n}^{*},\varepsilon_{\;\:m}^{ij}\boldsymbol{l}^{m}\right\rangle -k^{\left[ji\right]}-\tau^{\left(ji\right)}\right)-\frac{1}{2}\hat{n}\left[\,^{3}q^{ij}\right].
\end{align*}

The last equation is the tensor equation:
\begin{align*}
\,^{3}T^{ikj} & =\kappa\left(\frac{1}{3}\left(\mathcal{D}^{j}\right)\,^{3}q^{ik}-\frac{1}{12}\,^{3}q^{ij}\mathcal{D}^{k}+\left(\mathcal{\sigma}^{kij}+\varepsilon_{\;\:m}^{ij}l^{km}\right)\right).
\end{align*}
Splitting the index $\left(i,j\right)$ into its symmetric and antisymmetric
part gives: 
\begin{align}
0 & =\kappa\left(\frac{1}{3}\,^{3}q^{\left(i\right|k}\boldsymbol{\mathcal{D}}^{\left|j\right)}-\frac{1}{12}\,^{3}q^{ij}\mathcal{D}^{k}+\mathcal{\sigma}^{k\left(ij\right)}\right)\label{eq:con7}
\end{align}
\begin{align*}
\,^{3}T^{ikj} & =\kappa\left(\frac{1}{3}\,^{3}q^{\left[i\right|k}\boldsymbol{\mathcal{D}}^{\left|j\right]}+\varepsilon_{\;\:m}^{ij}l^{km}\right)
\end{align*}
The symmetric part gives another constraint to $\mathcal{H}$, while
the antisymmetric part gives solution to $\,^{3}T^{ikj}$.

Collecting all the constraints on $\mathcal{H}$, namely, (\ref{eq:con1}),
(\ref{eq:con2}), (\ref{eq:con3}), (\ref{eq:con4}), (\ref{eq:con5}),
(\ref{eq:con6}), and (\ref{eq:con7}), and solving them nicely gives:
\begin{equation}
\begin{array}{ccccc}
\left\langle \hat{n}^{*},\boldsymbol{\tau}^{i}\right\rangle =\frac{1}{6}\mathcal{D}^{i}, & \qquad & \left\langle n^{*},\textrm{tr}\boldsymbol{\sigma}\right\rangle \equiv-\frac{1}{4}\mathcal{D}\left(\hat{n}\right) & \qquad & \left\langle \hat{n}^{*},\boldsymbol{\sigma}^{\left(ij\right)}\right\rangle =-\frac{1}{12}\,^{3}q^{ij}\mathcal{D}\left(\hat{n}\right)\\
 & \,\\
\tau^{ji}=\frac{1}{6}\mathcal{D}\left(\hat{n}\right)\,^{3}q^{ij},\quad\tau^{\left[ji\right]}=0 &  & \left(\textrm{tr}\sigma\right)^{i}\equiv-\frac{1}{12}\mathcal{D}^{i}, &  & \mathcal{\sigma}^{k\left(ij\right)}=-\frac{1}{3}\,^{3}q^{\left(i\right|k}\boldsymbol{\mathcal{D}}^{\left|j\right)}+\frac{1}{12}\,^{3}q^{ij}\mathcal{D}^{k}
\end{array}\label{eq:consH}
\end{equation}
Finally, we write the additional variables in terms of $\mathcal{D}$:
\begin{align}
 & \Delta^{i}=\alpha^{i}=\kappa\left(\frac{1}{6}\mathcal{D}^{i}-\left\langle \hat{n}^{*},\boldsymbol{k}^{i}\right\rangle \right),\nonumber \\
 & \Delta^{ij}=\kappa\left(\frac{1}{2}\left\langle \hat{n}^{*},\varepsilon_{\;\:m}^{ij}\boldsymbol{l}^{m}\right\rangle -k^{\left[ji\right]}\right)-\frac{1}{2}\hat{n}\left[\,^{3}q^{ij}\right],\label{eq:metrix}\\
 & K^{ij}=\mathcal{K}^{ij}=\kappa\left(-\frac{1}{6}\mathcal{D}\left(\hat{n}\right)\,^{3}q^{ij}-\frac{1}{2}\left\langle \hat{n}^{*},\varepsilon_{\;\:m}^{ij}\boldsymbol{l}^{m}\right\rangle -k^{\left(ji\right)}\right)-\frac{1}{2}\hat{n}\left[\,^{3}q^{ij}\right],\nonumber \\
 & \,^{3}T^{ikj}=\kappa\left(\frac{1}{3}\,^{3}q^{\left[i\right|k}\mathcal{D}^{\left|j\right]}+\varepsilon_{\;\:m}^{ij}l^{km}\right),\nonumber 
\end{align}
while the remaining ones, namely $\Theta\left(\hat{n}\right),$ $\Theta^{i},$
$\,^{3}\nabla_{\hat{n}}\,^{3}q^{jk},$ and $\,^{3}\nabla_{i}\,^{3}q^{jk},$
are zero due to the metric compatibility (\ref{eq:metricomp}).

Equations in (\ref{eq:consH}) indicate that the hypermomentum $\mathcal{H}$
is constrained, one could choose $\mathcal{D}\left(\hat{n}\right)$
and $\mathcal{D}^{i}$ from the dilation $\boldsymbol{\mathcal{D}}$,
and $\boldsymbol{k}_{i}$, $\boldsymbol{l}^{k}$ from the rotation
$\Omega$ as the free quantities. The shear $\Lambda$ is determined
by $\boldsymbol{\mathcal{D}}$ and $\Omega$, and could be obtained
by inserting the constraint (\ref{eq:consH}) to (\ref{eq:shear1})-(\ref{eq:shear3}):
\begin{align}
\Lambda\left(\hat{n},\hat{n}\right) & =\textrm{tr}\boldsymbol{\sigma}=\frac{1}{4}\mathcal{D}\left(\hat{n}\right)\hat{n}-\frac{1}{12}\mathcal{D}^{i}\partial_{i},\nonumber \\
\Lambda\left(\hat{n},\partial_{i}\right) & =\boldsymbol{\tau}_{i}=\frac{1}{6}\left(-\mathcal{D}_{i}\hat{n}+\mathcal{D}\left(\hat{n}\right)\delta_{i}^{j}\partial_{j}\right),\label{eq:metrixon}\\
\Lambda\left(\partial_{i},\partial_{j}\right) & =\boldsymbol{\sigma}_{ij}=\frac{1}{12}\,^{3}q_{ij}\mathcal{D}\left(\hat{n}\right)\hat{n}+\left(-\frac{1}{3}\delta_{\left[i\right.}^{k}\boldsymbol{\mathcal{D}}_{\left.j\right]}+\frac{1}{12}\,^{3}q_{ij}\mathcal{D}^{k}\right)\partial_{k}.\nonumber 
\end{align}
Together with the metric-compatible (3+1) energy, momentum, and stress
equations (in normal coordinate) as follows \cite{nuaing}:
\begin{align}
 & \frac{1}{2}\left(\,^{3}\mathcal{R}-\mathrm{tr}\left(K^{2}\right)+\left(\mathrm{tr}K\right)^{2}\right)=\kappa E,\label{eq:sem1}\\
 & \frac{1}{2}\left(\,^{3}\nabla_{j}\left(K_{i}^{\,\:j}+\Delta_{\;i}^{j}\right)-\,^{3}\nabla_{i}K_{j}^{\,\:j}+\,^{3}T_{j\,\:i}^{\,\:\:k}K_{k}^{\,\:j}+\Delta_{i}K_{j}^{\,\:j}-\Delta^{j}K_{ji}-\hat{n}\left[\,^{3}\omega_{j\,\:i}^{\,\:\:j}\right]\right)=\kappa p_{i},\nonumber \\
 & \hat{n}\left[K_{\left(ij\right)}\right]-\,^{3}\nabla_{\left(i\right.}\Delta_{\left.j\right)}+\,^{3}R_{\left(ij\right)}+\left(\mathrm{tr}K\right)K_{\left(ij\right)}-K_{\left(i\right|}^{\,\:\:k}K_{k\left|j\right)}-K_{\left(i\right|k}\Delta_{\;\:\left|j\right)}^{k}=\kappa\mathcal{S}_{ij}-\frac{1}{2}\,^{3}q_{ij}\kappa\left(\mathcal{S}-E\right),\nonumber 
\end{align}
(\ref{eq:metrix}) and (\ref{eq:metrixon}) describe metric compatible
MAGR systems.

\subsubsection*{Special Case III: Torsionless Connection in Normal Coordinate}

For the last case, we apply the torsionless condition (\ref{eq:torsionless})
in the normal coordinate. With this, we have 6 independent variables:
$\Theta\left(\hat{n}\right),$ $^{3}\alpha$, $K$, $\,^{3}\omega$,
$\Theta_{i}=\Delta_{i},$ and $\Delta^{ji}=\mathcal{K}^{ij}$. Note
that the traceless torsion constraint (\ref{eq:EoM3}) is automatically
satisfied in this case, so we have 8 equation that must be satisfied,
consisting 1 scalar equation:
\begin{align}
\Delta_{\:\:i}^{i}-K_{i}^{\:\;i}=-\kappa\left(\frac{1}{4}\mathcal{D}\left(\hat{n}\right)+\left\langle \hat{n}^{*},\textrm{tr}\boldsymbol{\sigma}\right\rangle \right),\label{eq:scalar-1}
\end{align}
3 vectors equations:
\begin{align*}
 & \Delta^{i}-\alpha^{i}=\kappa\left(\left(\textrm{tr}\sigma\right)^{i}+\frac{1}{12}\mathcal{D}^{i}\right),\\
 & -\Delta^{i}+\,^{3}\nabla_{j}\,^{3}q^{ji}-\frac{1}{2}\,^{3}q_{jk}\,^{3}\nabla^{i}\,^{3}q^{jk}=\kappa\left(\frac{1}{3}\mathcal{D}^{i}-\left\langle \hat{n}^{*},\boldsymbol{\tau}^{i}+\boldsymbol{k}^{i}\right\rangle \right),\\
 & \frac{1}{2}\,^{3}q_{jk}\,^{3}\nabla_{i}\,^{3}q^{jk}-\Delta_{i}=\kappa\left\langle \hat{n}^{*},\boldsymbol{k}_{i}-\boldsymbol{\tau}_{i}\right\rangle ,
\end{align*}
3 matrix equations:
\begin{align*}
 & \left(\Theta\left(\hat{n}\right)+\frac{1}{2}\,^{3}q_{kl}\left(\hat{n}\left[\,^{3}q^{kl}\right]+\Delta^{kl}+\Delta^{lk}\right)\right)\,^{3}q^{ij}-\hat{n}\left[\,^{3}q^{ij}\right]-\Delta^{ij}-\Delta^{ji}=\kappa\left(\tau^{ji}+k^{ji}\right),\\
 & K_{i}^{\;j}-\Delta_{\;\:i}^{j}+\left(\Delta_{\;\:l}^{l}-K_{l}^{\;l}+\Theta\left(\hat{n}\right)-\frac{1}{2}\,^{3}q_{kl}\left(\hat{n}\left[\,^{3}q^{kl}\right]+\Delta^{kl}+\Delta^{lk}\right)\right)\delta_{i}^{j}=\kappa\left(-\frac{1}{3}\mathcal{D}\left(\hat{n}\right)\delta_{i}^{j}+\tau_{\;i}^{j}-k_{\;i}^{j}\right),\\
 & \Delta_{\;\:i}^{j}-K_{\;i}^{j}=\kappa\left(-\frac{1}{12}\delta_{i}^{j}\mathcal{D}\left(\hat{n}\right)-\,^{3}q^{jk}\left\langle \hat{n}^{*},\boldsymbol{\sigma}_{ik}+\varepsilon_{ikm}\boldsymbol{l}^{m}\right\rangle \right),
\end{align*}
and 1 tensor equation:
\begin{align*}
\left(\Delta^{j}-\alpha^{j}+\,^{3}\nabla_{l}\,^{3}q^{jl}-\Theta^{j}-\frac{1}{2}\,^{3}q_{lm}\,^{3}\nabla^{j}\,^{3}q^{lm}\right)\delta_{i}^{k}\\
+\left(\Theta_{i}+\frac{1}{2}\,^{3}q_{lm}\,^{3}\nabla_{i}\,^{3}q^{lm}\right)\,^{3}q^{jk}-\,^{3}\nabla_{i}\,^{3}q^{jk} & =\kappa\left(\frac{1}{3}\left(\mathcal{D}^{j}\right)\delta_{i}^{k}-\frac{1}{12}\delta_{i}^{j}\mathcal{D}^{k}+\,^{3}q^{jl}\left(\mathcal{\sigma}_{\:\:il}^{k}+\varepsilon_{ilm}l^{km}\right)\right).
\end{align*}
Here the variable $\Delta_{i}=\Theta_{i}$, $\Delta_{\;\:i}^{j}=\mathcal{K}_{i}^{\:\:j}$
by the torsionless condition.

The three vector equation could be simplified as follows:
\begin{align}
\Delta^{i}-\alpha^{i}= & \kappa\left(\left(\textrm{tr}\sigma\right)^{i}+\frac{1}{12}\mathcal{D}^{i}\right),\label{eq:ve1}\\
\,^{3}\nabla_{j}\,^{3}q^{ji}-2\Delta^{i}= & \kappa\left(\frac{1}{3}\mathcal{D}^{i}-2\left\langle \hat{n}^{*},\boldsymbol{\tau}^{i}\right\rangle \right),\label{eq:ve2}\\
\,^{3}\nabla_{j}\,^{3}q^{ji}-\,^{3}q_{jk}\,^{3}\nabla^{i}\,^{3}q^{jk}= & \kappa\left(\frac{1}{3}\mathcal{D}^{i}-2\left\langle \hat{n}^{*},\boldsymbol{k}^{i}\right\rangle \right).\label{eq:ve3}
\end{align}
By raising the indices $(i,j)$ the matrix equations could be written
as follows:
\begin{align*}
 & \left(\Theta\left(\hat{n}\right)+\frac{1}{2}\,^{3}q_{kl}\left(\hat{n}\left[\,^{3}q^{kl}\right]+\Delta^{kl}+\Delta^{lk}\right)\right)\,^{3}q^{ij}-\hat{n}\left[\,^{3}q^{ij}\right]-\Delta^{ij}-\Delta^{ji}=\kappa\left(\tau^{ji}+k^{ji}\right),\\
 & K^{ij}-\Delta^{ji}+\left(\Delta_{\;\:l}^{l}-K_{l}^{\;\:l}+\Theta\left(\hat{n}\right)-\frac{1}{2}\,^{3}q_{kl}\left(\hat{n}\left[\,^{3}q^{kl}\right]+\Delta^{kl}+\Delta^{lk}\right)\right)\,^{3}q^{ij}=\kappa\left(-\frac{1}{3}\mathcal{D}\left(\hat{n}\right)\,^{3}q^{ij}+\tau^{ji}-k^{ji}\right),\\
 & \Delta^{ji}-K^{ji}=\kappa\left(-\frac{1}{12}\,^{3}q^{ij}\mathcal{D}\left(\hat{n}\right)-\left\langle \hat{n}^{*},\boldsymbol{\sigma}^{ij}+\varepsilon_{\;\;m}^{ij}\boldsymbol{l}^{m}\right\rangle \right).
\end{align*}
These 3 equations needs to be splitted into their symmetric and antisymmetric
parts:
\begin{align}
\left(\Theta\left(\hat{n}\right)+\frac{1}{2}\,^{3}q_{kl}\left(\hat{n}\left[\,^{3}q^{kl}\right]+\Delta^{kl}+\Delta^{lk}\right)\right)\,^{3}q^{ij}-\hat{n}\left[\,^{3}q^{ij}\right]-2\Delta^{\left(ij\right)} & =\kappa\left(\tau^{\left(ji\right)}+k^{\left(ji\right)}\right),\label{eq:syma}\\
0 & =\kappa\left(\tau^{\left[ji\right]}+k^{\left[ji\right]}\right),\label{eq:antsyma}
\end{align}

\begin{align}
K^{ij}-\Delta^{\left[ji\right]}+\left(\Delta_{\;\:l}^{l}-K_{l}^{\;l}+\Theta\left(\hat{n}\right)-\frac{1}{2}\,^{3}q_{kl}\left(\hat{n}\left[\,^{3}q^{kl}\right]+\Delta^{kl}+\Delta^{lk}\right)\right)\,^{3}q^{ij} & =\kappa\left(-\frac{1}{3}\mathcal{D}\left(\hat{n}\right)\,^{3}q^{ij}+\tau^{\left(ji\right)}-k^{\left(ji\right)}\right),\label{eq:symb}\\
-\Delta^{\left[ji\right]} & =\kappa\left(\tau^{\left[ji\right]}-k^{\left[ji\right]}\right),\label{eq:antsymb}
\end{align}
and:
\begin{align}
\Delta^{\left(ji\right)}-K^{ji}= & \kappa\left(-\frac{1}{12}\,^{3}q^{ij}\mathcal{D}\left(\hat{n}\right)-\left\langle \hat{n}^{*},\boldsymbol{\sigma}^{\left(ij\right)}\right\rangle \right).\label{eq:symc}\\
\Delta^{\left[ji\right]}= & -\kappa\left\langle \hat{n}^{*},\varepsilon_{\;\;m}^{ij}\boldsymbol{l}^{m}\right\rangle .\label{eq:antsymc}
\end{align}
Notice that $K^{ij},$ by the torsionless condition (\ref{eq:torsionless}),
is symmetric on its indices. Subtituting (\ref{eq:antsyma}) to (\ref{eq:antsymb})
gives the antisymmetric part of $\Delta^{ij}$:
\begin{align}
\Delta^{\left[ij\right]}=-2\kappa\tau^{\left[ij\right].}\label{eq:con1a}
\end{align}
Subtituting (\ref{eq:antsymb}) with (\ref{eq:antsymc}) gives constraint
on $\mathcal{H}$:
\begin{align}
2\kappa\tau^{\left[ji\right]}= & \kappa\left\langle \hat{n}^{*},\varepsilon_{\;\;m}^{ij}\boldsymbol{l}^{m}\right\rangle ,\label{eq:con2a}
\end{align}
together with (\ref{eq:antsyma}). Equation (\ref{eq:symb}) and (\ref{eq:symc})
gives
\begin{align*}
\left(\Delta_{\;\:l}^{l}-K_{l}^{\;l}+\Theta\left(\hat{n}\right)-\frac{1}{2}\,^{3}q_{kl}\left(\hat{n}\left[\,^{3}q^{kl}\right]+\Delta^{kl}+\Delta^{lk}\right)\right)\,^{3}q^{ij}= & \kappa\left(-\frac{5}{12}\mathcal{D}\left(\hat{n}\right)\,^{3}q^{ij}+\tau^{\left(ji\right)}-k^{\left(ji\right)}-\left\langle \hat{n}^{*},\boldsymbol{\sigma}^{\left(ij\right)}\right\rangle \right)
\end{align*}
Contracting with $\,^{3}q_{ij}$ and substituting (\ref{eq:scalar-1}),
gives:
\begin{align}
\Theta\left(\hat{n}\right)-\frac{1}{2}\,^{3}q_{kl}\left(\hat{n}\left[\,^{3}q^{kl}\right]+\Delta^{kl}+\Delta^{lk}\right)= & \kappa\left(\frac{1}{3}\left(\mathrm{tr}\boldsymbol{\tau}-\mathrm{tr}\boldsymbol{k}\right)+\frac{2}{3}\left\langle \hat{n}^{*},\textrm{tr}\boldsymbol{\sigma}\right\rangle -\frac{1}{6}\mathcal{D}\left(\hat{n}\right)\right),\label{eq:bundar}
\end{align}
with $\mathrm{tr}\boldsymbol{\tau}=\,^{3}q_{ij}\tau^{ij}=\,^{3}q_{ij}\tau^{\left(ij\right)}$
and $\mathrm{tr}\boldsymbol{k}=\,^{3}q_{ij}k^{ij}=\,^{3}q_{ij}k^{\left(ij\right)}$.
Contracting (\ref{eq:syma}) with $\,^{3}q_{ij}$ gives:
\begin{align}
3\Theta\left(\hat{n}\right)+\frac{1}{2}\,^{3}q_{kl}\left(\hat{n}\left[\,^{3}q^{kl}\right]+\Delta^{kl}+\Delta^{lk}\right) & =\kappa\left(\mathrm{tr}\boldsymbol{\tau}+\mathrm{tr}\boldsymbol{k}\right).\label{eq:segitiga}
\end{align}
(\ref{eq:bundar}) and (\ref{eq:segitiga}) could be solved for $\Theta\left(\hat{n}\right)$
and $\,\nabla_{\hat{n}}\,^{3}q^{kl}$ as follows:
\begin{align}
 & \Theta\left(\hat{n}\right)=\kappa\left(\left(\frac{1}{3}\mathrm{tr}\boldsymbol{\tau}+\frac{1}{6}\mathrm{tr}\boldsymbol{k}\right)-\frac{1}{24}\mathcal{D}\left(\hat{n}\right)+\frac{1}{6}\left\langle \hat{n}^{*},\textrm{tr}\boldsymbol{\sigma}\right\rangle \right),\label{eq:bulath}\\
 & \left(\nabla_{\hat{n}}\,^{3}q^{kl}\right)=\frac{1}{3}\kappa\,^{3}q^{kl}\left(\mathrm{tr}\boldsymbol{k}+\frac{1}{4}\mathcal{D}\left(\hat{n}\right)-\left\langle \hat{n}^{*},\textrm{tr}\boldsymbol{\sigma}\right\rangle \right).\label{eq:deltan}
\end{align}
Inserting (\ref{eq:bulath}) and (\ref{eq:deltan}) to (\ref{eq:syma})
gives:
\begin{align}
\Delta^{\left(ij\right)}= & -\frac{1}{2}\kappa\left(\tau^{\left(ji\right)}+k^{\left(ji\right)}\right)-\frac{1}{2}\hat{n}\left[\,^{3}q^{ij}\right]+\frac{1}{2}\kappa\,^{3}q^{ij}\left(\frac{1}{3}\mathrm{tr}\boldsymbol{\tau}+\frac{2}{3}\mathrm{tr}\boldsymbol{k}+\frac{1}{12}\mathcal{D}\left(\hat{n}\right)-\frac{1}{3}\left\langle \hat{n}^{*},\textrm{tr}\boldsymbol{\sigma}\right\rangle \right),\label{eq:hmm}
\end{align}
while using (\ref{eq:hmm}) with (\ref{eq:con1a}) and (\ref{eq:symc})
gives:
\begin{align}
 & \Delta^{ij}=\kappa\left(-2\kappa\tau^{\left[ij\right]}-\frac{1}{2}\left(\tau^{\left(ji\right)}+k^{\left(ji\right)}\right)+\frac{1}{2}\,^{3}q^{ij}\left(\frac{1}{3}\mathrm{tr}\boldsymbol{\tau}+\frac{2}{3}\mathrm{tr}\boldsymbol{k}+\frac{1}{12}\mathcal{D}\left(\hat{n}\right)-\frac{1}{3}\left\langle n^{*},\textrm{tr}\boldsymbol{\sigma}\right\rangle \right)\right)-\frac{1}{2}\hat{n}\left[\,^{3}q^{ij}\right],\label{eq:segiti}\\
 & K^{ji}=K^{ij}=\kappa\left(\left\langle \hat{n}^{*},\boldsymbol{\sigma}^{\left(ij\right)}\right\rangle -\frac{1}{2}\left(\tau^{\left(ji\right)}+k^{\left(ji\right)}\right)+\,^{3}q^{ij}\left(\frac{1}{8}\mathcal{D}\left(\hat{n}\right)+\frac{1}{6}\mathrm{tr}\boldsymbol{\tau}+\frac{1}{3}\mathrm{tr}\boldsymbol{k}-\frac{1}{6}\left\langle \hat{n}^{*},\textrm{tr}\boldsymbol{\sigma}\right\rangle \right)\right)-\frac{1}{2}\hat{n}\left[\,^{3}q^{ij}\right].\label{eq:extcurves}
\end{align}

Finally, let us raise the index $i$ of the tensor equation, and then
contract it with $\,^{3}q_{jk}$ to obtain:
\begin{align*}
\Theta^{i}+\frac{1}{2}\,^{3}\nabla_{j}\,^{3}q^{ij} & =-\frac{1}{2}\left(\Delta^{i}-\alpha^{i}\right)+\kappa\left(\frac{1}{8}\mathcal{D}^{i}+\frac{1}{2}\,^{3}q^{in}\left(\mathcal{\sigma}_{\:\:nk}^{k}+\varepsilon_{nkm}l^{km}\right)\right).
\end{align*}
Inserting (\ref{eq:ve1}), and using torsionless condition, we obtain:
\begin{align}
\Delta^{i}+\frac{1}{2}\,^{3}\nabla_{j}\,^{3}q^{ij} & =\kappa\left(-\frac{1}{2}\left(\textrm{tr}\sigma\right)^{i}+\frac{1}{12}\mathcal{D}^{i}+\frac{1}{2}\,^{3}q^{in}\left(\mathcal{\sigma}_{\:\:nk}^{k}+\varepsilon_{nkm}l^{km}\right)\right).\label{eq:hmmm}
\end{align}
Using (\ref{eq:hmmm}) to (\ref{eq:ve2}) and (\ref{eq:ve3}), one
could solve for $\Delta^{i}$ and $\,^{3}\nabla_{i}\,^{3}q^{jk}$:
\begin{align}
 & \Delta^{i}=\kappa\left(\frac{1}{2}\left\langle \hat{n}^{*},\boldsymbol{\tau}^{i}\right\rangle -\frac{1}{24}\mathcal{D}^{i}-\frac{1}{4}\left(\textrm{tr}\sigma\right)^{i}+\frac{1}{4}\,^{3}q^{in}\left(\mathcal{\sigma}_{\:\:nk}^{k}+\varepsilon_{nkm}l^{km}\right)\right),\label{eq:segit}\\
 & \,^{3}\nabla_{i}\,^{3}q^{jk}=\frac{1}{3}\kappa\left(-\left\langle \hat{n}^{*},\boldsymbol{\tau}_{i}\right\rangle +2\left\langle \hat{n}^{*},\boldsymbol{k}_{i}\right\rangle -\frac{1}{12}\mathcal{D}_{i}-\frac{1}{2}\left(\textrm{tr}\sigma\right)_{i}+\frac{1}{2}\left(\mathcal{\sigma}_{\:\:in}^{n}+\varepsilon_{inm}l^{nm}\right)\right)\,^{3}q^{jk}.\label{eq:deltan2}
\end{align}
And last, from (\ref{eq:ve1}), we could obtain $\alpha^{i}$:
\begin{equation}
\alpha^{i}=\kappa\left(+\frac{1}{2}\left\langle \hat{n}^{*},\boldsymbol{\tau}^{i}\right\rangle -\frac{1}{8}\mathcal{D}^{i}-\frac{5}{4}\left(\textrm{tr}\sigma\right)^{i}+\frac{1}{4}\,^{3}q^{in}\left(\mathcal{\sigma}_{\:\:nk}^{k}+\varepsilon_{nkm}l^{km}\right)\right).\label{eq:accel}
\end{equation}

The solutions to the equation of motions, i.e., the equation of the
additional variables, are (\ref{eq:bulath}), (\ref{eq:deltan}),
(\ref{eq:segiti}), (\ref{eq:extcurves}), (\ref{eq:segit}), (\ref{eq:deltan2}),
and (\ref{eq:accel}), together with $\,^{3}T\left(\partial_{i},\partial_{j}\right)=0$.
Meanwhile, there are 2 constraints on $\mathcal{H}$, namely, (\ref{eq:antsyma})
and (\ref{eq:con2a}), are the constraint on the antisymmetric part
of $\tau^{ij}$ dan $k^{ij}$:
\begin{align}
\tau^{\left[ij\right]}= & -\frac{1}{2}\varepsilon_{\;\;m}^{ij}\left\langle \hat{n}^{*},\boldsymbol{l}^{m}\right\rangle ,\label{eq:cn1}\\
k^{\left[ij\right]}= & \frac{1}{2}\varepsilon_{\;\;m}^{ij}\left\langle \hat{n}^{*},\boldsymbol{l}^{m}\right\rangle .\label{eq:cn2}
\end{align}
Note that the index $i$ on $\tau^{ij}$ dan $k^{ij}$ define the
spatial indices, while the $j$ define the 'internal' indices. Only
the antisymmetric parts of $\tau^{ij}$ dan $k^{ij}$ are constrained.

Finally, using (\ref{eq:cn1})-(\ref{eq:cn2}), we could rewrite the
additional variables and collect all the results:
\[
\Theta\left(\hat{n}\right)=\kappa\left(\left(\frac{1}{3}\mathrm{tr}\boldsymbol{\tau}+\frac{1}{6}\mathrm{tr}\boldsymbol{k}\right)-\frac{1}{24}\mathcal{D}\left(\hat{n}\right)+\frac{1}{6}\left\langle \hat{n}^{*},\textrm{tr}\boldsymbol{\sigma}\right\rangle \right),
\]
\begin{align*}
 & \Theta^{i}=\Delta^{i}=\kappa\left(\frac{1}{2}\left\langle \hat{n}^{*},\boldsymbol{\tau}^{i}\right\rangle -\frac{1}{24}\mathcal{D}^{i}-\frac{1}{4}\left(\textrm{tr}\sigma\right)^{i}+\frac{1}{4}\,^{3}q^{in}\left(\mathcal{\sigma}_{\:\:nk}^{k}+\varepsilon_{nkm}l^{km}\right)\right),\\
 & \alpha^{i}=\kappa\left(+\frac{1}{2}\left\langle \hat{n}^{*},\boldsymbol{\tau}^{i}\right\rangle -\frac{1}{8}\mathcal{D}^{i}-\frac{5}{4}\left(\textrm{tr}\sigma\right)^{i}+\frac{1}{4}\,^{3}q^{in}\left(\mathcal{\sigma}_{\:\:nk}^{k}+\varepsilon_{nkm}l^{km}\right)\right),
\end{align*}
\begin{align*}
 & \left(\nabla_{\hat{n}}\,^{3}q^{kl}\right)=\frac{1}{3}\kappa\,^{3}q^{kl}\left(\mathrm{tr}\boldsymbol{k}+\frac{1}{4}\mathcal{D}\left(\hat{n}\right)-\left\langle \hat{n}^{*},\textrm{tr}\boldsymbol{\sigma}\right\rangle \right),\\
 & \,^{3}\nabla_{i}\,^{3}q^{jk}=\frac{1}{3}\kappa\left(\left\langle \hat{n}^{*},2\boldsymbol{k}_{i}-\boldsymbol{\tau}_{i}\right\rangle -\frac{1}{12}\mathcal{D}_{i}-\frac{1}{2}\left(\textrm{tr}\sigma\right)_{i}+\frac{1}{2}\left(\mathcal{\sigma}_{\:\:in}^{n}+\varepsilon_{inm}l^{nm}\right)\right)\,^{3}q^{jk}
\end{align*}
\begin{align*}
 & K^{ij}=K^{ji}=\kappa\left(\left\langle \hat{n}^{*},\boldsymbol{\sigma}^{\left(ij\right)}\right\rangle -\frac{1}{2}\left(\tau^{\left(ji\right)}+k^{\left(ji\right)}\right)+\,^{3}q^{ij}\left(\frac{1}{8}\mathcal{D}\left(\hat{n}\right)+\frac{1}{6}\mathrm{tr}\boldsymbol{\tau}+\frac{1}{3}\mathrm{tr}\boldsymbol{k}-\frac{1}{6}\left\langle \hat{n}^{*},\textrm{tr}\boldsymbol{\sigma}\right\rangle \right)\right)-\frac{1}{2}\hat{n}\left[\,^{3}q^{ij}\right].\\
 & \Delta^{ij}=\kappa\left(\varepsilon_{\;\;m}^{ij}\left\langle \hat{n}^{*},\boldsymbol{l}^{m}\right\rangle -\frac{1}{2}\left(\tau^{\left(ji\right)}+k^{\left(ji\right)}\right)+\frac{1}{2}\,^{3}q^{ij}\left(\frac{1}{3}\mathrm{tr}\boldsymbol{\tau}+\frac{2}{3}\mathrm{tr}\boldsymbol{k}+\frac{1}{12}\mathcal{D}\left(\hat{n}\right)-\frac{1}{3}\left\langle \hat{n}^{*},\textrm{tr}\boldsymbol{\sigma}\right\rangle \right)\right)-\frac{1}{2}\hat{n}\left[\,^{3}q^{ij}\right],
\end{align*}
with:
\begin{align*}
T_{i\;j}^{\;k} & =\omega_{i\;j}^{\;k}-\omega_{j\;i}^{\;k}=0.
\end{align*}
Together with the torsionless (3+1) energy, momentum, and stress equations
(in normal coordinate) \cite{nuaing}:
\begin{align}
 & \frac{1}{2}\left(\,^{3}\mathcal{R}-2\mathrm{tr}\left(K\Delta\right)+\left(\mathrm{tr}K\right)\left(\mathrm{tr}\Delta\right)-\mathrm{tr}\left(\Delta^{2}\right)-\hat{n}\left[\mathrm{tr}\Delta\right]\right.\nonumber \\
 & \left.\quad+\,^{3}q^{ij}\hat{n}\left[K_{ij}\right]+\,^{3}\nabla_{i}\alpha^{i}-\,^{3}q^{ij}\,^{3}\nabla_{i}\Theta_{j}-\Theta_{i}\left(\alpha^{i}+\Theta^{i}\right)+\Theta\left(\hat{n}\right)\left(\mathrm{tr}\Delta+\mathrm{tr}K\right)\right)=\kappa E,\label{eq:sem2}\\
 & \frac{1}{2}\left(2\,^{3}\nabla_{j}\Delta_{\,\:i}^{j}-\,^{3}\nabla_{i}\Delta_{\,\:j}^{j}+2\Theta_{i}\Delta_{\,\:j}^{j}-2\alpha^{j}K_{ij}-\partial_{i}\Theta\left(\hat{n}\right)+\hat{n}\left[\Theta_{i}\right]-\hat{n}\left[\,^{3}\omega_{j\,\:i}^{\,\:\:j}\right]\right)=\kappa p_{i},\nonumber \\
 & \hat{n}\left[K_{ij}\right]-\Theta_{i}\Theta_{j}-\,^{3}\nabla_{\left(i\right.}\Theta_{\left.j\right)}+\,^{3}R_{\left(ij\right)}+\left(\mathrm{tr}\Delta+\Theta\left(\hat{n}\right)\right)K_{ij}-2K_{\left(i\right|k}\Delta_{\;\:\left|j\right)}^{k}=\kappa\mathcal{S}_{ij}-\frac{1}{2}\,^{3}q_{ij}\kappa\left(\mathcal{S}-E\right),\nonumber 
\end{align}
we have describe torsionless MAGR systems.

\section{Discussions and Conclusions}

\subsubsection*{The Relations between Stress-Energy and Hypermomentum Tensor}

As mentioned clearly in \cite{Iosifidis}, the stress-energy tensor
$\mathcal{T}$ is not completely independent from the hypermometum
tensor $\mathcal{H}$. This is because they originate from the same
source $S_{\mathrm{matter}}$. The relation comes from the fact that
$\delta_{\omega}\delta_{g}S_{\mathrm{matter}}=\delta_{g}\delta_{\omega}S_{\mathrm{matter}}.$
With definitions (\ref{eq:tiga}) and (\ref{eq:ggg}), one could derive
\cite{Iosifidis}:
\begin{equation}
\frac{\partial\mathcal{T}_{\mu\nu}}{\partial\omega_{\alpha\,\:\beta}^{\:\,\:\gamma}}=\frac{\partial\mathcal{H}_{\:\:\gamma}^{\alpha\,\:\beta}}{\partial g^{\mu\nu}}-\frac{1}{2}g_{\mu\nu}\mathcal{H}_{\:\:\gamma}^{\alpha\,\:\beta}=\frac{1}{\sqrt{-\mathfrak{g}}}\frac{\partial\left(\sqrt{-\mathfrak{g}}\mathcal{H}_{\:\:\gamma}^{\alpha\,\:\beta}\right)}{\partial g^{\mu\nu}},\label{eq:constraii}
\end{equation}
which is a relation that must be satisfied by both $\mathcal{T}$
and $\mathcal{H}$. For the zero hypermomentum case, $\mathcal{H}_{\:\:\gamma}^{\alpha\,\:\beta}=0$,
and therefore, the stress-energy tensor needs to satisfy $\frac{\partial\mathcal{T}_{\mu\nu}}{\partial\omega_{\alpha\,\:\beta}^{\:\,\:\gamma}}=0$,
i.e., $\mathcal{T}$ is independent from the connection $\omega$.
For the next two cases, i.e., the metric or torsionless cases, the
hypermomentum are constrained (equation (\ref{eq:metrixon}) for metric
and (\ref{eq:cn1})-(\ref{eq:cn2}) for torsionless case). These will
restrict the degrees of freedom of the corresponding stress-energy
tensor (in the set of equations (\ref{eq:sem1}) for metric and (\ref{eq:sem2})
for torsionless case), via relation (\ref{eq:constraii}). One could
apply the (3+1) formulation to relation (\ref{eq:constraii}), but
this is out of the scope in this article.

\subsubsection*{Comments on the Symmetricity Constraint}

Without the symmetricity constraint (\ref{eq:ELmafr4}), the Euler-Lagrange
equation (\ref{eq:ELmafr1}) will become:
\begin{equation}
f'\left(\mathcal{R}\right)R_{\alpha\beta}-\frac{1}{2}\left(f\left(\mathcal{R}\right)+\chi^{\mu}T_{\lambda\,\:\mu}^{\:\:\lambda}\right)g_{\alpha\beta}=\mathcal{\kappa T}_{\alpha\beta}.\label{eq:palsu}
\end{equation}
Even with reducing the degrees of freedom of the general affine connection
$\omega$ via the traceless torsion constraint (\ref{eq:ELmaffr3}),
the Ricci tensor $R_{\alpha\beta}$ will still possess the antisymmetric
part:
\begin{align}
R_{\left[\alpha\beta\right]} & =\frac{1}{2}\left(R_{\alpha\beta}-R_{\beta\alpha}\right)=\frac{1}{2}\left(\nabla_{\mu}T_{\alpha\,\:\beta}^{\:\:\mu}+\nabla_{\alpha}T_{\beta\,\:\mu}^{\:\:\mu}+\nabla_{\beta}T_{\mu\,\:\alpha}^{\:\:\mu}-R_{\alpha\beta\,\:\,\mu}^{\:\:\:\:\:\mu}-T_{\alpha\,\:\beta}^{\:\:\sigma}T_{\mu\,\:\sigma}^{\:\:\mu}\right).\label{eq:antisym}
\end{align}
Inserting the traceless torsion constraint (\ref{eq:ELmaffr3}) to
(\ref{eq:antisym}) will only give:
\begin{align}
R_{\left[\alpha\beta\right]} & =\frac{1}{2}\left(\nabla_{\mu}T_{\alpha\,\:\beta}^{\:\:\mu}-R_{\alpha\beta\,\:\,\mu}^{\:\:\:\:\:\mu}\right),\label{eq:antisym-1}
\end{align}
where $R_{\alpha\beta\,\:\,\mu}^{\:\:\:\:\:\mu}$ is the homothetic
curvature that arise from the non-metricity condition \cite{Iosifidis3,Jimenez2,Iosifidis4}.
Since $g_{\alpha\beta}$ and $\mathcal{T}_{\alpha\beta}$ is symmetric,
from (\ref{eq:palsu}), we obtain another condition from the dynamics,
relating the torsion and the non-metricity as follows:
\[
\nabla_{\mu}T_{\alpha\,\:\beta}^{\:\:\mu}=R_{\alpha\beta\,\:\,\mu}^{\:\:\:\:\:\mu},
\]
This unnecessary constraint could be avoided by introducing the symmetricity
constraint (\ref{eq:ELmafr4}) in the level of the action, such that
the antisymmetric part $R_{\left[\alpha\beta\right]}$ simply does
not enter the equation of motion, see Section III.

\subsubsection*{Conclusions and Further Remarks}

Let us summarize the results achieved from our work. First, we have
performed the (3+1)-decomposition to the torsion tensor and the non-metricity
factor. These are equations (\ref{eq:T00})-(\ref{eq:Tij}) and (\ref{eq:q1})-(\ref{eq:q4})
in Section II, which are written using the additional variables (see
\cite{nuaing}). Second, we have given a complete derivation of the
equation of motion of Metric-Affine $f\left(\mathcal{R}\right)$-gravity,
i.e., (\ref{eq:ELmafr1-1}), (\ref{eq:ELmafr2-1}), and (\ref{eq:ELmaffr3-1});
these equations are well-known, and we do not add some new results,
except a different way to rederive the hypermomentum equation and
the explicit use of the symmetricity constraint. Third, following
\cite{Hehl}, we decomposed the hypermomentum tensor into the dilation,
shear, and rotational (spin) parts, and then split each part into
its temporal and spatial components. Hence, we have 10 hypersurface
variables that are responsible for the construction of the hypermomentum
tensor. The main result of this article is the (3+1)-formulation of
the hypermomentum equation, together with the (3+1) traceless torsion
constraint. They are obtained by inserting the (3+1) torsion tensor,
(3+1) non-metricity factor, and (3+1) hypermomentum tensor from the
previous sections to equation (\ref{eq:EoM2})-(\ref{eq:EoM3}). These
resulting equations are given at the end of Section IV, containing
2 scalar equations, 4 vector equations, 3 matrix equations, and 1
tensor equation of order $\binom{2}{1}$. We provided several special
cases for the (3+1) hypermomentum equation, namely, the zero hypermomentum,
the metric connection, and the torsionless connection case in Section
V.

Collecting the results in this article with our previous results in
\cite{nuaing}, we have obtained a complete (3+1)-formulation for
Metric-Affine General Relativity, or Generalized Palatini Gravity.
The first difference of this theory with the standard GR lies in the
connection: in MAGR, the connection is not constrained. This difference
manifests in the (3+1) equation of motion. Similar to the standard
GR, in the MAGR, the covariant stress-energy-momentum equation is
decomposed into 3 canonical equations: the generalized energy (or
Hamiltonian) equation, the generalized momentum (or the diffeomorphism)
equation, and the generalized stress-energy equation. However, unlike
the (3+1) standard GR, none of the (3+1) MAGR are constraint equations
since all of them contain the derivative of quantities with respect
to the time coordinate. The second main difference between MAGR and
standard GR is the dependence of matter on the connection, which in
MAGR, will give rise to the hypermomentum tensor. In the standard
Palatini gravity, this term vanished, and with the traceless torsion
constraint, it gives equivalent dynamics to the standard GR. In our
works, we have shown that for vanishing hypermomentum, the (3+1) hypermomentum
equation, together with traceless torsion constraint, are equivalent
to the metric compatibility and torsionless condition, hence constraining
the affine connection to be Levi-Civita. Furthermore, with the metric
compatibility and torsionless condition, one could retrieve the Hamiltonian
and diffeomorphism constraint from the generalized energy and momentum
equation, hence obtaining the standard dynamics of GR.

As mentioned in \cite{nuaing}, there are several subtleties we encounter
along with our work. The first subtlety is technical, concerning the
tracelessness of the torsion tensor, which is given by hand, or in
other words, is introduced at the kinematical level via the projective
transformation. First, taking the tracelessness of the torsion tensor
is not the only way to solve the projective invariance problem; one
could equally choose the Weyl vector to be zero, or more generally,
the linear combination of both the trace of torsion and the latter,
as discussed in \cite{Iosifidis,Iosifidis4}. The choice will affect
the form of the resulting (3+1) equations. From the theoretical perspective,
it is important to obtain the (3+1) formulation in the most general
setting. Second, it is more favorable to introduce the tracelessness
of torsion (or moreover, the vanishing of Weyl vector) at the dynamical
level, where for the traceless torsion case, this had been done in
\cite{Percacci} by adding extra terms in the action, at least quadratic
in torsion. Another more phenomenological subtlety is on the physical
interpretation of the hypermomentum tensor. There had been ongoing
research on the candidates of matter that will give a non-vanishing
hypermomentum, and hence causing the torsion and non-metricity in
spacetime. Some progress on these problem had been addressed in \cite{exp1,exp2,exp3,exp4,exp5,delhom3,delhom4,ariki,damian3,damian4}
and research in this direction is highly encouraged. We expect our
works in this article could contribute to the further research in
the direction of the complete Hamiltonian analysis, the existence
of the Cauchy problem, some extension to (3+1) Metric-Affine\textit{-}$f\left(\mathcal{R}\right)$
and other Metric-Affine theories.

\section*{Acknowledgement}

A. S.  was funded by Riset Institut Teknologi Bandung. F. P. Z. was
funded by Riset Institut Teknologi Bandung and RISTEKDIKTI.

\end{document}